\documentclass[11pt]{LMA}

\newcommand{\shn}{MLRSM}
\newcommand{\lgn}{mediated left-right supersymmetric model}

\begin{document}
\title{\sffamily Left-Right Supersymmetry as the Origin of Flavor Physics
\vspace{.5em}}

\author[1]{Sean Fraser\thanks{sean.fraser@kbfi.ee}}

\author[1,2,3]{Emidio Gabrielli\thanks{emidio.gabrielli@cern.ch}}

\author[1]{Carlo Marzo\thanks{carlo.marzo@kbfi.ee}}

\author[1]{Luca Marzola\thanks{luca.marzola@cern.ch}}

\author[1]{Martti Raidal\thanks{martti.raidal@cern.ch}}

\affil[1]{\KBFI}

\affil[2]{Physics Department, University of Trieste, Strada Costiera 11, 34151 Trieste, Italy and INFN, Sezione di Trieste, Via Valerio 2, 34127 Trieste, Italy}

\affil[3]{IFPU - Institute for Fundamental Physics of the Universe, Via Beirut 2, 34151 Trieste, Italy}

\date{\today\vspace{-0.3cm}}

\maketitle
\begin{abstract}  
\noindent We propose a new framework for the observed flavor hierarchy and mixing based on left-right supersymmetry. The model contains the most minimal Higgs sector
consisting only of gauge doublets which forbid the standard model Yukawa couplings. New mediator fields then connect the left- and right-chiral fermion sectors and result in effective tree level Yukawa couplings for the third generation charged fermions. The remaining fermions, including all neutrinos, acquire effective Yukawa couplings sourced by the supersymmetry breaking sector at loop level. We predict new TeV range scalars, as well as heavier fermions and vector bosons, that can be discovered at the LHC and future colliders.
\end{abstract}  

\vspace{1cm}

\section{Introduction} 
\label{sec:Introduction}
Despite the phenomenological success of the standard model (SM), the mass and mixing hierarchies in the quark and lepton sectors lack of a proper understanding. 
Specifically, the inferred SM Yukawa couplings of 
fermions to the Higgs field span many orders of magnitude, and 
their numerical values do not reflect any underlying principle.
The properties of the discovered Higgs boson~\cite{Higgs}
support the existence of these interactions, although 
current collider experiments only probe the Yukawa couplings of the third generation charged fermions~\cite{YukawaCMS,YukawaATLAS}.

A possible approach toward understanding the flavor puzzle postulates that SM Yukawa couplings are low-energy effective interactions induced by new dynamics. Clearly, the radiative generation of these quantities would provide a simple understanding for the relative lightness of the first two SM generations. However, at the same time, all radiative generation models generally struggle with the top quark Yukawa coupling, as the observed value of this parameter seemingly requires generation mechanisms borderline with perturbative unitarity. The problem of the top quark is then typically addressed through external tree-level contributions, which inevitably impair the attractiveness of the framework. Furthermore, from a model-building perspective, it is also desirable that the Yukawa couplings of bottom quark and tau lepton also result from the same dynamics as the top quark Yukawa coupling. 

Motivated by these considerations, we show here that the mentioned shortcomings of radiative generation models can be easily avoided once two well-known symmetries are considered: supersymmetry (SUSY) and left-right (LR) gauge symmetry. In particular, we show that the corresponding interactions \textit{result} in effective Yukawa couplings generated at the tree level for charged fermions of the third SM generation, and at higher orders in perturbation theory for the remaining particles.   

Irrespectively of SUSY, the left-right gauge group is well-motivated as a natural solution to the observed parity violation~\cite{Mohapatra:1974gc, Mohapatra:1974hk,Senjanovic:1975rk, Mohapatra:1977mj,Senjanovic:1978ev} and to the strong $CP$-problem~\cite{Mohapatra:1978fy,Mohapatra:1997su,Diaz-Cruz:2016pmm}. On top of that, left-right models have also been extensively used to radiatively generate the mass spectrum of the lightest SM fermions~\cite{Weinberg:1972ws, Georgi:1972hy, Mohapatra:1974wk, Balakrishna:1987qd, Balakrishna:1988ks,Gabrielli:2016vbb}.  
An open question in this line of model building concerns the gauge structure of the Higgs sector, which usually relies on the bidoublet representation~\cite{Mohapatra:1974gc, Mohapatra:1974hk,Senjanovic:1975rk, Mohapatra:1977mj,Senjanovic:1978ev,Duka:1999uc,Tello:2010am,Maiezza:2010ic} rather than on  doublet~\cite{Davidson:1987mh,Brahmachari:2003wv} scalar representations. 

Although the former choice dominates the literature, the latter possibility has gained renewed interest~\cite{Gabrielli:2016vbb} within frameworks for the radiative generation of the fermion mass hierarchy, as it automatically forbids the presence of all tree-level SM Yukawa couplings. In these schemes, fermion masses are then recovered by considering three generations, $i=1,2,3,$ of vector-like fermions $U_i, D_i, E_i$ that mediate between the SM fermions of opposite chirality and the corresponding Higgs bosons, implementing the so-called universal see-saw mechanism~\cite{Davidson:1987mh,Brahmachari:2003wv,Berezhiani:1983hm,Berezhiani:1985in,Rajpoot:1987fca,Berezhiani:1991ds,Berezhiani:1995yk,Siringo:2004hm}. An alternative approach developed in Ref.~\cite{Gabrielli:2016vbb} uses higher dimensional operators, generated at the one-loop level by the dynamics of a dark sector, to reproduce the full fermion mass hierarchy. The dark sector consists of a set of new fermions, singlet under the SM gauge group but charged under a new $U(1)_F$ gauge symmetry, and of messenger fields charged under all interactions. The latter, which carry the same quantum number as the squarks and sleptons of supersymmetric models, transmit the flavor structure and chiral symmetry breaking from the dark sector to the SM via new scalar-fermion interactions that resemble the gaugino interactions of supersymmetric models. 

Continuing this line of model building, in the present paper we abandon the universal aspect of the mentioned seesaw models and propose a different paradigm where, instead of a dark sector, the interactions that result in the radiative generation of SM Yukawa couplings arise from supersymmetry. Concretely, adopting once again the doublet representation of Higgs scalars in a left-right framework to forbid the SM Yukawa interactions, we consider a minimal scenario where the chiral sectors of the theory are joined by a single copy of each mediator $U,D,E$. As anticipated, only the charged fermions of one of the SM generations consequently acquire tree-level interactions that reduce to the corresponding Yukawa couplings via the universal seesaw mechanism below the $SU(2)_R$ breaking scale. The mass terms of the remaining fermions, instead, arise from loop-level interactions that are fully determined by the supersymmetric structure of the theory. The soft breaking sector and the Majorana gaugino masses, in particular, respectively provide the necessary departure from flavor universality and chiral symmetry breaking. The renormalizability of the model ensures that finite, and therefore predictive, results are obtained for the Yukawa couplings of the lightest generations\footnote{The fact that the SUSY interactions alone produce finite contribution to the SM Yukawa couplings was first acknowledged in Refs.~\cite{Banks:1987iu,Borzumati:1999sp} within the context of the minimal supersymmetric model.}. In this way, the proposed interplay between the left-right symmetry and the supersymmetry delineates a new supersymmetric framework to address the flavor puzzle, which we name the \textit{\lgn{}} (\shn). 

As a first test for the \shn, we tackle the hierarchy that characterizes the quark sector of the SM and the observed flavor mixing, parameterized in the Cabibbo-Kobayashi-Maskawa (CKM) matrix. After computing the tree- and loop-level contributions entering the quark mass matrices generated by the new interactions, we show through a dedicated numerical analysis that the \shn{} is able to match the observed quark mass pattern and mixing in a natural way.
We also sketch the leptonic sector of the model, where the hierarchy of the charged lepton masses is explained by the same mechanism behind the generation of the quark hierarchy. Differently, the omission of a `neutrino mediator', which would transform trivially under the gauge group of the theory, prevents the generation of neutrino masses both at the tree and one-loop level. These are then generated only the two-loop level, resulting in Dirac neutrinos that are naturally lighter than the rest of the SM fermions. We argue that the peculiar origin of neutrino masses sources here the different mixing pattern observed in the leptonic sector.   
 
The paper is structured as follows: in section~\ref{sec:Model} we introduce the superpotential and particle content defining the \shn, and delineate the general properties of the framework. Our investigation of the SM flavor hierarchy begins in section~\ref{sec:Qs}, where we show how third generation charged fermions acquire their masses through the interaction with the new mediator sector. The mechanisms behind the generation of the remaining quark and charged lepton mass terms are detailed in section~\ref{sec:Ofcb}, whereas in section~\ref{sec:FH} we present a numerical analysis that demonstrates how the SM quark sector flavor structure is faithfully reproduced. Finally, our conclusions are presented in section~\ref{sec:Conclusions}.

\section{The \shn} 
\label{sec:Model}
We propose a new supersymmetric scheme based on a realization of the left-right symmetry that involves only Higgs doublets. The SM Yukawa operators are then forbidden by gauge invariance and the chiral sectors of the theory must be necessarily bridged via a new mediator sector. The particle content of the theory is then shared amongst: 
\begin{itemize}
	\item Left- and right-chiral sectors, which host the quark and lepton doublets of the corresponding chirality, including right-handed neutrinos, the relative $SU(2)_{L/R}$ gauge bosons and two Higgs fields each. The setup of a chiral sector resembles that of left-handed fermions and Higgs doublets in the minimal supersymmetric standard model (MSSM).
	\item A `mediator sector' that bridges the two chiral sectors of the theory. The mediators transform as singlets under the $SU(2)_L \times SU(2)_R$ group but possess color and hypercharge -- or, better, $B-L$ charge -- and interact with the remaining fields via gauge interactions, as well as via fundamental Yukawa couplings that involve the fermion and Higgs doublets of either chirality. In absence of mediators and symmetry breaking, which induces the mixing of neutral gauge bosons, the two chiral sectors are completely decoupled.
\end{itemize}
A comprehensive list of the considered superfields with their charges and gauge multiplicities is presented in Table~\ref{Qcharges}.

As mentioned in the introduction, our setup resemble that of universal seesaw models where the SM Yukawa couplings emerge from a double seesaw mechanism after heavy vector-fermion mediators are integrated out \cite{Brahmachari:2003wv,Davidson:1987mh,Rajpoot:1987fca,Siringo:2004hm,Berezhiani:1985in,Berezhiani:1983hm,Berezhiani:1995yk,Berezhiani:1991ds}. In our construction, however, we abandon universality and consider a minimal setup for the mediator sector which contains only one generation of each kind of mediator (super)field. This simplification gives rise to a series of implication that we briefly sketch below: 
\begin{itemize}
	\item  All SM Yukawa couplings arise from higher dimensional interaction terms generated, after the spontaneous breaking of $SU(2)_R$, by integrating out the involved mediator fields and superpartners. The 4-dimensional SM Yukawa operators then cease to exist as local operators above a scale of validity of the emerging effective theory, with non trivial implications on the renormalization group  running of the coupling constants and scenarios of grand-unification. 
	\item The considered mediator sector implements the universal seesaw mechanism only for one generation of quarks and charged leptons. These particles, which we identify with the third generation of the SM, are the only fermions that consequently acquire tree-level mass terms. 
	\item The Yukawa interactions resulting in the mass terms for 1st and 2nd generation charged fermions are generated only at the one-loop level. These particles are consequently naturally lighter than the 3rd generation ones and the SM flavor hierarchy is recovered together with the emergence of an approximate $U(2)$ symmetry involving the lightest quark generations. The chiral symmetry-breaking terms that allow for the required loop diagrams are naturally provided by the gaugino masses. 
	\item No neutrino acquires mass at the tree or one loop-level. The absence of a `neutrino mediator', which transforms as a singlet under the full gauge group of the theory, forces Dirac masses to appear only at the two-loop level, in agreement with the the observed fermion mass hierarchy. 
	\item	The different structures of the PMNS and CKM matrices can be potentially related to the different origin of neutrino masses. 
\end{itemize} 

\begin{table}[h]
	\centering
	\begin{tabular}{c|c|c}
		\toprule
		Chiral Superfield & $U(1)_{B-L}\times SU(2)_L\times SU(2)_R\times SU(3)_c$ & Particle content\\
		\midrule
		\rowcolor[gray]{.95}
		Left-handed sector: & & \\
		& & \\
$Q_i = (u_i,d_i)$  & (1/3, 2, 1, 3) &    Left-handed (s)quarks \\
$L_i = (\nu_i, \ell_i)$  & (-1, 2, 1, 1) &     Left-handed (s)leptons\\
$H_u = (\phi_u^+, \phi_u^0)$& (1, 2, 1, 1) &      $SU(2)_L$ up-type Higgs(ino) \\
$H_d = (\phi_d^0, \phi_d^-)$& (-1, 2, 1, 1) &     $SU(2)_L$ down-type Higgs(ino)\\
		& & \\
		\rowcolor[gray]{.95}
		Mediator sector:& & \\
		& & \\
$U$  & (4/3, 1, 1, $3$)  &  Up-type (s)mediator\\
$D$  & (-2/3, 1, 1 , $ 3$) &  Down-type (s)mediator\\
$E$  & (-2, 1, 1, 1)  & Charged lepton (s)mediator\\
$\bar U$  & (-4/3, 1, 1, $\bar3$)  &  Up-type (s)mediator\\
$\bar D$  & (2/3, 1, 1 , $\bar3$) &  Down-type (s)mediator\\
$\bar E$  & ( 2, 1, 1, 1)  & Charged lepton (s)mediator\\
		& & \\
		\rowcolor[gray]{.95}
        Right-handed sector: & &  \\
		& & \\
$\bar{Q}_i = (\bar{d}_i,\bar{u}_i)$  & (-1/3, 2, 1, $\bar 3$) &    Right-handed (s)quarks \\
$\bar{L}_i = (\bar{\ell}_i, \bar{\nu}_i)$  & (1, 2, 1, 1) &     Right-handed (s)leptons\\
$\bar{H}_u = (\bar{\phi}_u^0, \bar{\phi}_u^-)$& (-1, 2, 1, 1) &      $SU(2)_R$ up-type Higgs(ino) \\
$\bar{H}_d = (\bar{\phi}_u^0, \bar{\phi}_u^-)$ & (1, 2, 1, 1) &     $SU(2)_R$ down-type Higgs(ino) \\
		\bottomrule
	\end{tabular}
 \caption{Particle content of the \shn{} specified in terms of left-handed chiral superfields. We define the $B-L$ charge in a way that $Q = I_3 + (B-L)/2$.}
 	\label{Qcharges}
\end{table}

The only mass terms allowed by this setup in the superpotential, before SUSY and left-right symmetry breaking, are the mediator masses and two $\mu$ terms for the Higgs supermultiplets. Explicitly, we have:
\begin{align}
\label{eq:suppot}
\mathcal{W} = &
\mathcal{Y}_u^{i} \, Q_i H_u \, \bar{U} 
- \bar{\mathcal{Y}}_u^{i} \,\bar{ Q}_i \bar H_u  \,U
-  \mathcal{Y}_d^{i} \, Q_i H_d \, \bar{D} 
+ \bar{\mathcal{Y}}_d^{i} \, \bar Q_i \bar H_d \, D 
- \mathcal{Y}_e^i \, L_i H_d \bar E  
+ \bar{\mathcal{Y}}_e^{i} \,  \bar L_i \bar H_d  E  
\\\nn&\qquad+
{\color{blue}b_L L H_u   + \bar{b}_L  \bar H_u \bar L 
+
t_L L L \bar{E} + \bar{t}_L E \bar L \bar L }
+ {\color{ForestGreen}
s_D Q L \bar{D} + \bar{s}_D D \bar L\bar Q}  
\\ \nn &\qquad +{\color{ForestGreen}
\kappa_Q QQ D + \bar{\kappa}_Q \bar Q \bar Q\bar D }
+
{\color{orange}
\kappa U D \bar{E}+\bar{\kappa}  E\bar{U} \bar{D} 
}
\\ \nn &\qquad +
\mu_U U \bar{U} + \mu_D D \bar{D} + \mu_EE\bar{E}+ \mu_L H_u H_d + \mu_R \bar H_u \bar H_d\,.
\end{align}
Here, the first and the last line contain the interactions that are needed to generate the effective SM Yukawa couplings below the $SU(2)_R$ breaking and the mediator scales, as well as the $\mu$-terms for the four Higgs doublets and the mediators. Terms in blue are analogous to the so-called bilinear and trilinear $R$-parity violating terms of the MSSM, the phenomenology of which has been extensively studied in literature (see for instance Ref.~\cite{Abada:2000xr} and references therein).
Terms in green, instead, are responsible for the proton decay through diagrams as the one shown in Fig.~\ref{fig:Fig_prot_dec}. Forbidding either of these vertices is sufficient to ensure the proton stability.
Finally, terms in orange include interactions amongst the mediators which could  induce sizable contributions to proton decay in part of the parameter space once the $\kappa_Q$ vertices are allowed.

Additional terms $\mathcal{W} \supset \kappa_D U D D + \bar \kappa_D \bar{D}\bar{D}\bar{U}$ can be included in the superpotential within extensions of the model containing more generations of mediators, but vanish identically in the present case. Notice also that the Higgs doublets carry the same quantum number as lepton doublets. It could then be possible to modify the present framework by including different generations of Higgs doublets in place of lepton doublets, extending the superpotential to terms as $\mathcal{W}\supset +\kappa_u H_u H_u E + \bar \kappa_u \bar E \bar H_u \bar H_u  + \kappa_d  H_d H_d \bar E + \bar \kappa_d E \bar H_d \bar H_d$ which necessarily involve different generations of Higgs doublets. Leptons would then be identified with the charginos and neutralinos of the theory, but in the present analysis we do not pursue this fascinating possibility.
 
\subsection{Avoiding proton decay} 
\label{sub:Avoiding proton decay}

\begin{figure}[h]
  \centering
    \includegraphics[width=.6\textwidth]{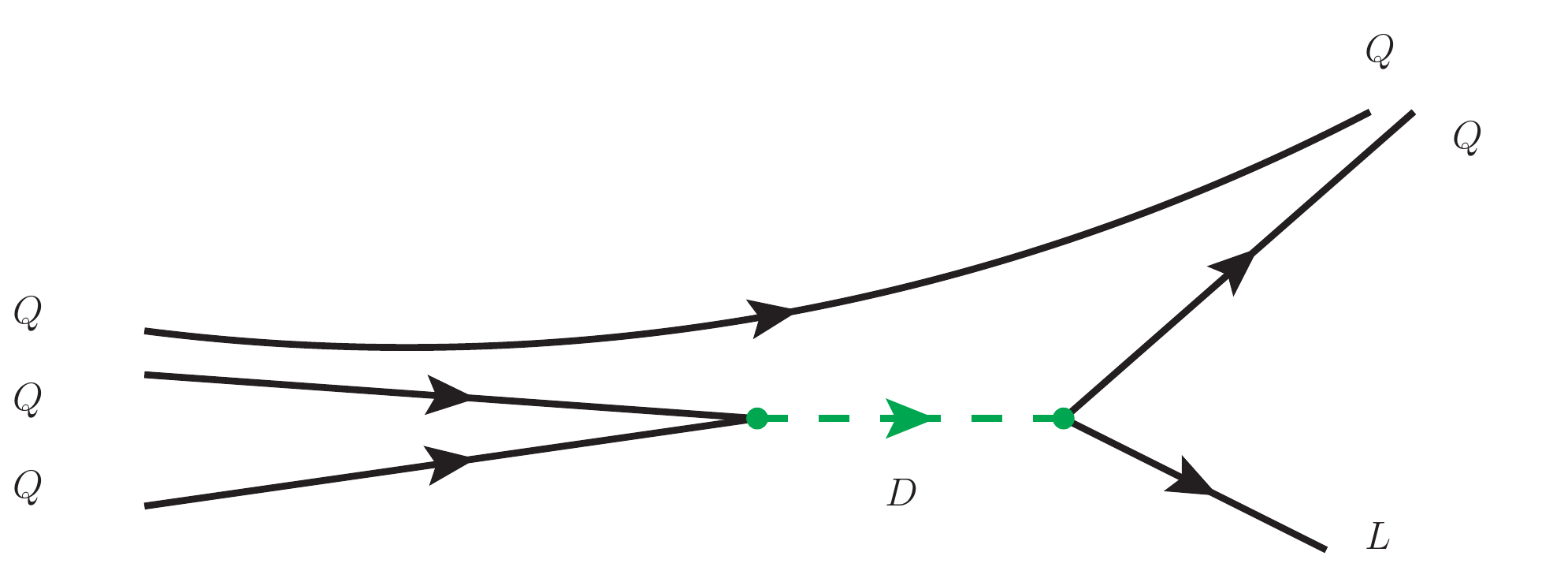}
  \caption{An example of proton decay diagram induced by the new mediators through the introductions rendered in green in eq.~\eqref{eq:suppot}.}
  \label{fig:Fig_prot_dec}
\end{figure}

It is possible to impose discrete symmetries to forbid superpotential terms that induce diagrams as the one presented in Fig.~\ref{fig:Fig_prot_dec}, resulting in the decay of protons.

For instance, introducing a $\mathbb{Z}_2$ symmetry that plays here the role of the $R$ parity of the MSSM, we find the following viable textures:

\begin{table}[htb!]
	\centering
	\begin{tabular}{c|c|c|c|c|c|c|c|c|c|c|c|c|c|c}
		\toprule
		Chiral Superfield: 
		& $Q$  & $L$  & $H_u$& $H_d$& $U$  &$D$  &$E$  &$\bar U$  &
		$\bar D$  & $\bar E$  & $\bar Q$  &$\bar L$  &$\bar{H}_u$&$\bar{H}_d$\\
		\midrule
		Texture I: &
		 $-$ &  $-$ & + &  + &  $-$ & $-$ &  $-$ & $-$  & $-$ &$-$  & $-$ &  $-$ &  + & +  \\
		 \hline
		Texture II: &
		 +  &  + & $-$ &  $-$ &  $-$ & $-$ &  $-$ & $-$  & $-$ &$-$  & + &  + &  $-$ & $-$  \\
		 \hline
		 Texture III: &
		 +  &  $-$ & + &  $-$ &  + & $-$ &  + & +  & $-$ &+  & + &  $-$ &  + & $-$  \\
		 \hline
		 Texture IV: &
		 $-$  &  $-$ & $-$ &  + &  + & $-$ &  $-$ & +  & $-$ &$-$  &$-$ &  $-$ &  $-$ & +  \\
		\bottomrule
	\end{tabular}
	\caption{Examples of $\mathbb{Z}_2$ parity textures which prevent the proton decay.}
	\label{tab:z2charges}
\end{table}
   
The textures I and II forbid all the terms in the second and third line of eq.~\eqref{eq:suppot}, whereas III forbids only terms in the latter. Texture IV, instead, allows for the terms analogous to the bilinear $R$-parity violating terms of the MSSM (the first two rendered in blue) and the mediator interactions (in orange), on top of the terms in the first and last lines. 
\vspace{1cm}
\par

In order to perform a first exploration of the framework, in this paper we focus on the quark interactions and show how the \shn{} explains the SM flavor hierarchy. To this purpose, we will consider only the following subset of terms present in the superpotential
\begin{eqnarray} \label{superpot}
\mathcal{W}_{Quarks} &=& + \mathcal{Y}_u^{i}\, Q_i\,H_u \,\bar{U} - \bar{\mathcal{Y}}_u^{i} \,\bar{Q}_{i} \bar{H}_u \,U - \mathcal{Y}_d^{i}\,Q_i\,H_d \,\bar{D} +
\bar{\mathcal{Y}}_d^{i} \,\bar{Q}_{i} \bar{H}_d \,D + \mu_U U \bar U +\, \mu_D D \bar D \, ,
\end{eqnarray}
which are allowed by every texture identified above. The corresponding soft-breaking sector then comprises
\begin{eqnarray} \label{supersoft}
\mathcal{L}_{Soft} &=& - \left(\mathcal{T}_u^i \tilde{Q}_i\,H_u\,\tilde{\bar{U}} - \bar{\mathcal{T}}_u^i\, \tilde{\bar{Q}}_i\, \bar{H}_u\,\tilde{U} 
-\mathcal{T}_d^i \tilde{Q}_i\,H_d\,\tilde{\bar{D}} + \bar{\mathcal{T}}_d^i\, \tilde{\bar{Q}}_i\, \bar{H}_d\,\tilde{D}
+\, b_{\mu_U} \tilde{U} \tilde{\bar U} + b_{\mu_D} \tilde{D} \tilde{\bar D} + \text{c.c.}\right) + \nn \\
&& - \,\tilde{Q}^{\dagger}_i m^2_{Q}{}^{i,\,i} \tilde{Q}_i -  \tilde{\bar{Q}}^{\dagger}_{i} m^2_{\bar{Q}}{}^{i,\,i} \tilde{\bar{Q}}_i - 
m^2_{U} \tilde{U}^* \tilde{U} -  m^2_{\bar{U}} \tilde{\bar{U}}^* \tilde{\bar{U}} - 
m^2_{D} \tilde{D}^* \tilde{D} -  m^2_{\bar{D}} \tilde{\bar{D}}^* \tilde{\bar{D}} + \nn \\
&& -\, M_{\tilde{g}}\,\tilde{g}\,\tilde{g} - M_{\tilde{W}_L}\,\tilde{W}_L\,\tilde{W}_L - M_{\tilde{W}_R}\,\tilde{W}_R\,\tilde{W}_R - M_{\tilde{B}}\,\tilde{B}\,\tilde{B}\,
+ \dots \,,
\end{eqnarray}
where we left understood the $SU(2)$ doublet-doublet contractions. Analogously, the suspension points stands for contribution that do not involve color interactions and therefore do not enter the present analysis. Without loss of generality, we take diagonal squark mass matrices on the gauge interaction basis. 

In order to reduce the number of free parameters and increase symmetry, during the following we will often invoke a mirror $\mathbb{Z}_2$ symmetry between the left and right sectors of the theory. To further simplify the study, we also parametrize the VEVs induced by the spontaneous symmetry breaking as follows: 
\begin{eqnarray}
<\phi_u^0> = \frac{v_u}{\sqrt{2}} = \frac{v_{H}}{\sqrt{2}} \sin{\theta}\,, && <\phi_d^0> = \frac{v_d}{\sqrt{2}}  = \frac{v_{H}}{\sqrt{2}} \cos{\theta}\,,   \nn \\
<\bar{\phi}_u^0> = \frac{v_{\bar{u}}}{\sqrt{2}} = \frac{v_{\bar{H}}}{\sqrt{2}} \sin{\bar{\theta}}\,, && <\bar{\phi}_d^0> = \frac{v_{\bar{d}}}{\sqrt{2}} = \frac{v_{\bar{H}}}{\sqrt{2}} \cos{\bar{\theta}}\,.
\end{eqnarray}
The full list of Feynman rules corresponding to the terms in eq.~\eqref{superpot} and \eqref{supersoft} is presented in Appendix~\ref{FRUp}. These will be used to build the one-loop Green functions entering the low energy effective operators that characterize the SM limit of the theory.

\subsection{Gauge sector}
\label{sec:Gs}

The tree-level masses for the $W_L$ vector boson and its right counterpart $W_R$ are straightforwardly given by 
\begin{eqnarray} 
	\label{VEVs}
v_H = \frac{2 M_{W_L}}{g_{2}}\,,\,\,\, v_{\bar{H}} = \frac{2 M_{W_R}}{g_{2}}\, ,   
\end{eqnarray}
where we have highlighted the relation with the Higgs VEVs and imposed a $\mathbb{Z}_2$ mirror symmetry so that $g_2 = g_{2L} = g_{2R}$. The analysis of the neutral sector is more involved due to the presence of the extra Abelian  $U(1)_{B-L}$ group. The masses of the $Z$ and $Z'$ states are derived from the matrix
\begin{eqnarray}
\mathcal{M}_{B} \equiv \left(\begin{array}{ccc}
	g_1^2\left(v_H^2 +   v_{\bar{H}}^2\right)^2 & - \frac{g_1\,g_{2L}}{2}\, v_{H}^2 & - \frac{g_1\,g_{2R}}{2}\, v_{\bar{H}}^2  \\
	- \frac{g_1\,g_{2L}}{2}\,v_H^2 & \frac{g_{2L}^2}{4}\, v_{H}^2 & 0  \\
	- \frac{g_1\,g_{2R}}{2}\,v_{\bar{H}}^2  & 0 & \frac{g_{2R}^2}{4}\, v_{\bar{H}}^2 \\
\end{array}\right)\,. 
\end{eqnarray}
Forcing again the mirror symmetry, eq.~\eqref{VEVs} and the limit $v_H \ll v_{\bar{H}}$ imposed by the present Large Hadron Collider (LHC) searches~\cite{Sirunyan:2018pom,Aaboud:2018spl,CMS:2016abv,Aaboud:2017buh}, we obtain
\begin{eqnarray} \label{g2mZp2}
g^2_2 = 4\,g^2_1\,\frac{2\,M^2_{W_L} - M_Z^2}{M_Z^2 - M^2_{W_L}} \,, \,\,
M^2_{Z'} = \frac{M^2_{W_L}\,M_{W_R}^2}{2\,M^2_{W_L} - M_Z^2} \,,
\end{eqnarray}
leaving two free parameters that we take to be $M_{W_R}$ and $g_1$.

Notice that eq.~\eqref{g2mZp2} and the imposed mirror symmetry set the ratio of 
the $B-L$ and $SU(2)$ gauge couplings, as well the $Z'/W_{R}$ mass ratio, to:
\begin{eqnarray}
g_2 / g_1 \simeq 3 \,,\,\,\,\,\, M_{Z'} / M_{W_R} \simeq 0.8 \, .
\end{eqnarray}

\section{The third generation masses} 
\label{sec:Qs}

With the Lagrangian given on the basis where the squark mass matrices are diagonal, we define the following chiral multiplets 
\begin{eqnarray}
&& \xi_{L_u}^T = \left(u_1, u_2, u_3, U\right)\, ,\,\, \xi_{\bar{R}_u}^T = \left(\bar{u}_1, \bar{u}_2, \bar{u}_3, \bar{U}\right) \, , \nn \\ 
&& \xi_{L_d}^T = \left(d_1, d_2, d_3, D\right)\, ,\,\, \xi_{\bar{R}_d}^T = \left(\bar{d}_1, \bar{d}_2, \bar{d}_3, \bar{D}\right)\,. \nn
\end{eqnarray}
The spontaneous symmetry breaking of the two chiral of $SU(2)$ symmetries then induces tree-level left-right bilinear terms of the form 
\begin{eqnarray} \label{LagQ}
&&\mathcal{L}^2 = - \xi_{L_u}^T\, \mathcal{M}_{u}\, \xi_{\bar{R}_u} - \xi_{L_d}^T\, \mathcal{M}_{d}\, \xi_{\bar{R}_d} \,,
\end{eqnarray}
with an analogous term for the charged leptons.

At the tree-level, the \shn{} quark sector is consequently characterized by a $4\times 4$ matrix texture of the form 
\begin{eqnarray} \label{MuMd}
\mathcal{M}_{u} \equiv \left(\begin{array}{cc}
	{\bf 0}_{3\times 3} & \left( v_u/\sqrt{2} \right) \vec{\mathcal{Y}}_u\\
	\left(v_{\bar{u}}/\sqrt{2}\right) \vec{\bar{\mathcal{Y}}}_u &  \mu_U \\
\end{array}\right)\,,\quad 
\mathcal{M}_{d} \equiv \left(\begin{array}{cc}
	{\bf 0}_{3\times 3} & \left( v_d/\sqrt{2} \right) \vec{\mathcal{Y}}_d\\
	\left(v_{\bar{d}}/\sqrt{2}\right) \vec{\bar{\mathcal{Y}}}_d &  \mu_D \\
\end{array}\right)\,.
\end{eqnarray}
As these matrices have rank 2 regardless of the values of the mediator Yukawa couplings, they each admit two vanishing eigenvalues. 

In order to see this explicitly for the case of up-type quarks, we momentarily rotate eq.~\eqref{LagQ} to the corresponding mass eigenstates basis ${\xi'}_{L_u}$ -- the cases of down-type quarks and charged leptons follow in complete analogy. We then introduce two different unitary matrices, $Z_{L_u}$ and $Z_{\bar{R}_u}$, such that
\begin{eqnarray} \label{mix}
&&{\xi'}_{L_u} = Z_{L_u} \xi_{L_u} \,,\,\,\, {\xi'}_{\bar{R}_u} = Z_{\bar{R}_u} \xi_{\bar{R}_u} ,\nn \\
&& \xi_{L_u}^T\, \mathcal{M}_{u}\, \xi_{\bar{R}_u}  =   {\xi'}_{L_u}^T\,Z^*_{L_u}\, \mathcal{M}_{u}\,Z^{\dagger}_{\bar{R}_u}\,{\xi'}_{\bar{R}_u}
=   {\xi'}_{L_u}^T\, \mathcal{M}^D_{u}\,{\xi'}_{\bar{R}_u},
\end{eqnarray}
where $\mathcal{M}^D_{u}$ is the diagonal matrix containing the mass eigenvalues given by $\sqrt{\mathcal{M}_u\,\mathcal{M}_u^{\dagger}}$.

Going into the details of the procedure, with simple geometric considerations we determine the intermediate rotations that extract from the $\mathcal{Y}_u^{i}\, Q_i$ and $\bar{\mathcal{Y}}_u^{i}\, \bar{Q}_i$ terms
the only states that interact with the Higgs bosons of the two chiral sectors at the tree-level\footnote{Notice that the \shn{} introduces in each chiral sector a Yukawa \emph{vector} rather than a matrix.}, and that mixes the messenger $U$ and $\bar{U}$. By using the following angular parameterization 
\begin{eqnarray}
&& \vec{\mathcal{Y}}_{u} = \mathcal{Y}_{u} \left(\sin{\theta_{L_u}}, \cos{\phi_{L_u}} \cos{\theta_{L_u}}, \sin{\phi_{L_u}} \cos{\theta_{L_u}}\right)\,, \nn \\
&&  \vec{\bar{\mathcal{Y}}}_{u} = \mathcal{\bar{ Y}}_{u} \left(\sin{\theta_{\bar{R}_u}}, \cos{\phi_{\bar{R}_u}} \cos{\theta_{\bar{R}_u}}, \sin{\phi_{\bar{R}_u}} \cos{\theta_{\bar{R}_u}} \right)\,, 
\end{eqnarray}
the intermediate transformation we seek are easily found by means of two $O(2)$ rotations that align the starting bases $\xi_{L_u}$ and $\xi_{R_u}$   
to the direction of the corresponding Yukawa vectors. A possible choice is therefore:
\begin{eqnarray} \label{OLU}
\mathcal{O}_{L_u} &\equiv& \left(\begin{array}{cccc}
	\cos{\phi_{L_u}} & - \sin{\phi_{L_u}} \sin{\theta_{L_u}} & -  \cos{\phi_{L_u}}  \sin{\theta_{L_u}} & 0 \\
	0 & \cos{\phi_{L_u}} & - \sin{\phi_{L_u}} & 0 \\
	\sin{\theta_{L_u}} &  \sin{\phi_{L_u}} \cos{\theta_{L_u}} & \cos{\phi_{L_u}} \cos{\theta_{L_u}} & 0 \\
	0 & 0 & 0 & 1 \\
\end{array}\right) \,, 
\end{eqnarray}
with $\mathcal{O}_{\bar{R}_u} $ given by the same expression through the substitutions $\phi_{L_u}\to\phi_{\bar{R}_u}$ and $\theta_{L_u}\to\theta_{\bar{R}_u}$.

By acting with these matrices on eq.~\eqref{MuMd} we find 
\begin{eqnarray} \label{Rot1}
\mathcal{O}_{L_u}\, \mathcal{M}_{u}\,\mathcal{O}_{\bar{R}_u}^{T}  = 
\left(\begin{array}{cccc}
	0 & 0 & 0 & 0 \\
	0 & 0 & 0 &  0 \\
	0 & 0 & 0 &  \left(v_u/\sqrt{2}\right) \mathcal{Y}_{u} \\
	0 & 0 & \left(v_{\bar{u}}/\sqrt{2}\right) \mathcal{\bar{Y}}_{u} & \mu_U \\
\end{array}\right)  \,, 
\end{eqnarray}
and the the $4-$dimensional spaces spanned by $Q_i,\,U$ and $\bar{Q}_i,\,\bar{U}$ consequently split each into in two separate sectors. On one hand we have the appearance of two massive states, which we identify with the third generation up-type quark and the relative mediator field. On the other, the states `rotated out' of the interactions with the Higgs bosons form a chiral sector characterized at this stage by a $U(2)$ symmetry, remnant of the original $U(3)$ flavor symmetry.

The same procedure can also be applied to the down-type quarks and charged lepton, which consequently present similar mass spectrums. Having performed the intermediate rotation in eq.~\eqref{OLU}, it is now easy to derive the explicit form of the matrices entering the biunitary transformation $Z^*_{L_u}\, \mathcal{M}_{u}\,Z^{\dagger}_{\bar{R}_u}$ by separately diagonalizing 
\begin{eqnarray}
Z^*_{L_u}\, \mathcal{M}_{u}\,\mathcal{M}_{u}^{\dagger}\,Z^T_{L_u} = \left(\mathcal{M}_{u}^D\right)^2\, , \nn \\
Z_{\bar{R}_u}\,\mathcal{M}_{u}^{\dagger}\,\mathcal{M}_{u}\,Z^{\dagger}_{\bar{R}_u} = \left(\mathcal{M}_{u}^D\right)^2\, .
\end{eqnarray}
The resulting two non-zero mass eigenvalues have the form  
\begin{eqnarray} \label{eigMEx}
M^2_{u\,4,3} = \frac{1}{2} \left(\mathcal{K}_{L_u}^2 + \mathcal{K}_{\bar{R}_u}^2 + \mu_U^2 \pm \sqrt{\left(\mathcal{K}_{L_u}^2 + \mathcal{K}_{\bar{R}_u}^2 + \mu_U^2\right)^2 - 4\,\mathcal{K}_{L_u}^2\,\mathcal{K}_{\bar{R}_u}^2} \right) ,
\end{eqnarray}
where $\mathcal{K}_{L_u} = v_u/\sqrt{2} \, \mathcal{Y}_{u}$ and  $\mathcal{K}_{\bar{R}_u} = v_{\bar{u}}/\sqrt{2}\, \mathcal{\bar Y}_{u}$.

Working under the natural assumption that the electroweak scale be much below the $SU(2)_R$ breaking scale and the supersymmetric 
messenger one, $\mu_U$, the above masses take a very simple form
\begin{eqnarray}\label{eigM}
M^2_{u\,4} \simeq \mu_U^2 + \bar{\mathcal{K}}_{u}^2 + O\left( \frac{\mathcal{K}_{u}^2}{\mu_U^2 + \bar{\mathcal{K}}_{u}^2}\right)  \,,\,\, M^2_{u\,3} \simeq \bar{\mathcal{K}}_{u}^2 \, \frac{\mathcal{K}_{u}^2}{\mu_U^2 + \bar{\mathcal{K}}_{u}^2}\,+ O\left( \left(\frac{\mathcal{K}_{u}^2}{\mu_U^2 + \bar{\mathcal{K}}_{u}^2}\right)^2\right)\,,
\end{eqnarray}
whereas the mixing matrices in the same limit are
\begin{eqnarray}  \label{ZLU}
Z_{L_u} =   \text{diagonal phases} \cdot
\left(\begin{array}{cccc}
	1 & 0 & 0 & 0 \\
	0 & 1 & 0 &  0 \\
	0 & 0 & 1 & -\frac{\mathcal{K}_{u}\,\mu_U}{\mu_U^2 + \bar{\mathcal{K}}_{u}^2} \\
	0 & 0 & \frac{\mathcal{K}_{u}\,\mu_U}{\mu_U^2 + \bar{\mathcal{K}}_{u}^2} & 1 \\
\end{array}\right) \cdot  \mathcal{O}_{L_u}  + O\left(\frac{\mathcal{K}_{u}^2}{\mu_U^2 + \bar{\mathcal{K}}_{u}^2}\right) \, ,
\end{eqnarray}
and
\begin{eqnarray} \label{ZRU}
Z_{\bar{R}_u} =  \text{diagonal phases} \cdot
\left(\begin{array}{cccc}
	1 & 0 & 0 & 0 \\
	0 & 1 & 0 &  0 \\
	0 & 0 & -\frac{\mu_U}{\sqrt{\mu_U^2 + \bar{\mathcal{K}}_{u}^2}} & \frac{\bar{\mathcal{K}}_{u}}{\sqrt{\mu_U^2 + \bar{\mathcal{K}}_{u}^2}} \\
	0 & 0 & \frac{\bar{\mathcal{K}}_{u}}{\sqrt{\mu_U^2 + \bar{\mathcal{K}}_{u}^2}} & \frac{\mu_U}{\sqrt{\mu_U^2 + \bar{\mathcal{K}}_{u}^2}} \\
\end{array}\right)  \cdot  \mathcal{O}_{\bar{R}_u} + O\left( \frac{\mathcal{K}_{u}^2}{\mu_U^2 + \bar{\mathcal{K}}_{u}^2}\right)  \, .
\end{eqnarray}
The matrix factorization presented in eq.~\eqref{ZLU} and \eqref{ZRU}   highlights the terms responsible for the mixing between the 3rd generation SM gauge eigenstates and the extra messengers.  

If the left-right mirror symmetry was also imposed in the Yukawa sector by requiring $\mathcal{Y}_{u} = \mathcal{\bar Y}_{u}$, and similarly for $\mathcal{Y}_d$,
the expressions in  eq.~\eqref{eigMEx} would reduce to\footnote{In this case the messenger Yukawa matrices become Hermitian, as expected for an exact left-right symmetry. }
\begin{eqnarray} \label{rel1}
\mathcal{Y}^2_{u} = \frac{2\,M_{u\,3}\,M_{u\,4}}{v_{H}\,v_{\bar{H}}\,\sin \theta\,\sin \bar{\theta}} \,,\,\, 
\mathcal{Y}^2_{d} = \frac{2\,M_{d\,3}\,M_{d\,4}}{v_{H}\,v_{\bar{H}}\,\cos \theta\,\cos \bar{\theta}} \,,
\end{eqnarray}
and define the scale of the messenger
\begin{eqnarray} \label{rel2}
\mu_U^2 &=& M_{u\,4}^2 + M_{u\,3}^2 -  \bar{\mathcal{K}}_{u}^2 + \mathcal{K}_{u}^2 \simeq M_{u\,4}^2 -
\frac{M_{u\,3}\,M_{u\,4}}{\sin \theta\,\cos \bar{\theta}} \,\frac{v_{\bar{H}}}{v_H}\, ,\nn \\
\mu_D^2 &=& M_{d\,4}^2 + M_{d\,3}^2 - \left(\bar{\mathcal{K}}_{d}^2 + \mathcal{K}_{d}^2\right) \simeq M_{d\,4}^2 -
\frac{M_{d\,3}\,M_{d\,4}}{\cos \theta\,\sin \bar{\theta}} \,\frac{v_{\bar{H}}}{v_H}\, .
\end{eqnarray}

To conclude the section, we remark on a peculiarity of the model rooted in the left-right symmetric structure possessed by the Yukawa sector even in absence of the mirror symmetry. The mixing between the left chiral sector and the messenger, regulated by eq.~\eqref{ZLU} can be arbitrarily suppressed by simply imposing that $v_H$  be the lightest scale, regardless of the hierarchy between $v_{\bar{H}}$ and $\mu_U$. On the contrary, eq.~\eqref{ZRU} seems to support a maximally mixed state between the right-handed component of the third generation quark and the messenger field unless $v_{\bar{H}} \ll \mu_U$. Such a condition, however, is ruled out in our model by the perturbativity of $\mathcal{Y}_u$.  

This can be explicitly shown by expanding the exact solutions for the mass eigenvalues, eq.~\eqref{eigMEx}, in the limit $\mu_U \gg v_{\bar{H}}$. The resulting expressions maintains the same functional dependence on the masses of eq.~\eqref{rel1} but in this case $M_{u\,4} \simeq \mu_U \gg v_{\bar{H}}$. Because top quark mass measurements force $M_{u\,3} \simeq v_H/\sqrt{2}$ we find that $\mathcal{Y}_{u} > M_{u/,4}/v_{\bar{H}}$, leading to a  non-perturbative value of the coupling under the assumption $\mu_U \gg v_{\bar{H}},v_{H}$.

\section{The first two generations} 
\label{sec:Ofcb}
In the previous section we have shown that the \shn{} effectively prevents two of the three SM quark -- and charged leptons -- generations from interacting with the Higgs bosons at the tree-level. The breaking of left-right gauge symmetry defines a subspace in the fields space characterized by a $U(2)$ flavor symmetry, which must be broken in order to generate the masses of the involved SM particles. We will now show that the required dynamics is provided by the soft SUSY breaking terms and the same mediator Yukawa interactions, which fully determine the effective operators emerging at the loop-level in the model. 

To illustrate the dominant one-loop contributions, as well as the role of the soft parameters involved, we focus again on the up-quark case and explore the radiative origin of masses for the lightest two generation quarks. We seek loop corrections that result in the effective operator 
\begin{eqnarray} \label{EffYuk}
\mathcal{L}_{eff} = \frac{\mathcal{Y}_{eff}}{\Lambda_{eff}}\,\left(\bar{\psi}_L\,H\right) \left(\psi_R\,\bar{H}\right)\,,
\end{eqnarray}
which radiatively implements the universal see-saw mechanism for the light SM generations. Considering the interactions derived from the soft Lagrangian in eq.~\eqref{superpot} and listed in Appendix \ref{FRUp}, we see that the helicity flip induced by the Majorana gaugino, on top of the (s)messenger mediation, provide different ways to construct the operator. 

\subsection{The role of mediators}
With the Lagrangian given on the flavor eigenstates basis where the squark mass matrices are diagonal, and barring for the moment soft trilinear terms, the Yukawa interactions of the mediator itself must provide the required contact with the squarks and quarks of the first two SM generations.

The structure of the light quark mass matrix is then shaped by the mediator Yukawa interactions and by the (diagonal) squark mass matrices $m^2_{\tilde{Q}}$. The latter, in particular, play an important role in breaking the mass degeneracy of the light quarks. In more detail, one-loop corrections to the mass matrix in eq.~\eqref{Rot1} result in
\begin{eqnarray} \label{RotLoop}
\mathcal{O}_{L_u}\, \left(\mathcal{M}_{u} + \text{loop correction} \right)\,\mathcal{O}_{\bar{R}_u}^{T} \simeq
\left(\begin{array}{cc}
	\mathcal{M}_u &  \left(v_u/\sqrt{2}\right) \mathcal{Y}_{u} \\
	\left(v_{\bar{u}}/\sqrt{2}\right) \mathcal{\bar Y}_{u} & \mu_U \\
\end{array}\right)  \,,
\end{eqnarray}
and the diagrammatic expression of the $3\times3$ sub-matrix $\mathcal{M}_u$ containing the masses of light quarks can be given as
\begin{equation}
	\label{diags}
	\begin{tabular}{m{1cm} m{6cm}}
		$(\mathcal{M}_u^{\mu^2})_{ij} \quad =$ &\hspace{1cm} \includegraphics[width=\linewidth]{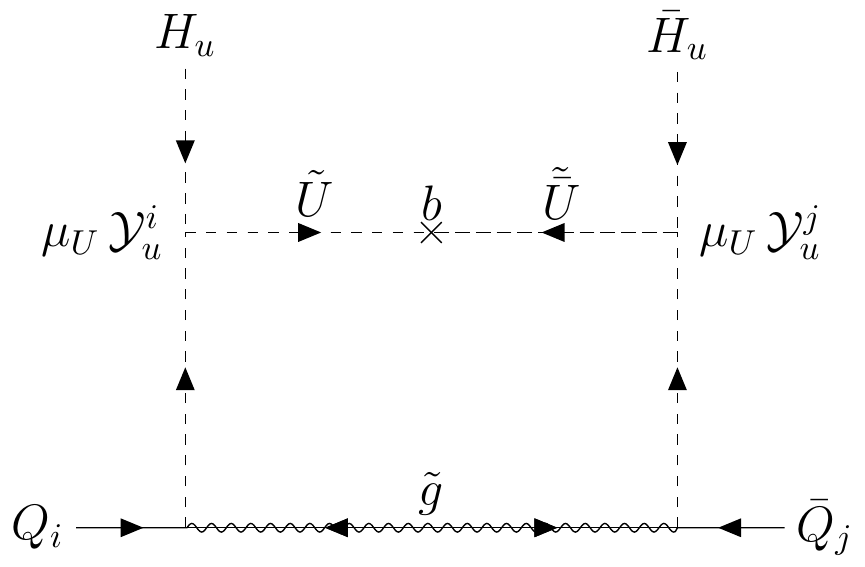}
	\end{tabular}	
	\end{equation}
which results in the matrix element
\begin{equation} \label{first}
(\mathcal{M}_u^{\mu^2})_{ij} =\sum_{\kappa=1}^2
\int \frac{d^4 k}{(2\pi)^4}\,\left(2\,C_{F}\,g_3^2\,\mu_U^2\right)\,M_{\tilde{g}}\,P_{\tilde{g}} \nn \\
\left( \mathcal{Y}_u^{i}\, P_{\tilde{Q}_{i}}\,\right) \left(  V^{\dagger}_{2 \kappa}\,P_{\tilde{u}_{\kappa}}\,V_{\kappa 1} \right)
\left( \mathcal{Y}_u^{j}\, P_{\tilde{\bar{Q}}_{j}}\,\right)\, ,
\end{equation}
where $C_F$ is the colour factor and the factors $P_{\tilde{f}}=P_{\tilde{f}}(k, M_{\tilde{f}}) = 1/(k^2 -M^2_{f})$ are due to the fermion propagators. We remark that these matrix elements are computed for vanishing external momenta.

\subsection{The role of trilinear terms}

Alternatively, quarks and squarks of the first two SM generation can be linked to the Higgs fields via trilinear terms from the soft SUSY breaking sector. This results in independent contributions to the light quark mass matrices provided that, in flavor space, the trilinear term be not aligned to the tree-level Yukawa vectors. The relevant interactions then result in a $3\times3$ sub-matrix diagrammatically written as  
\begin{equation} \label{diags2}
\begin{tabular}{m{1cm} m{6cm}}
	$(\mathcal{M}_u^{\mathcal{T}^2})_{ij} \quad =$ &\hspace{1cm} \includegraphics[width=\linewidth]{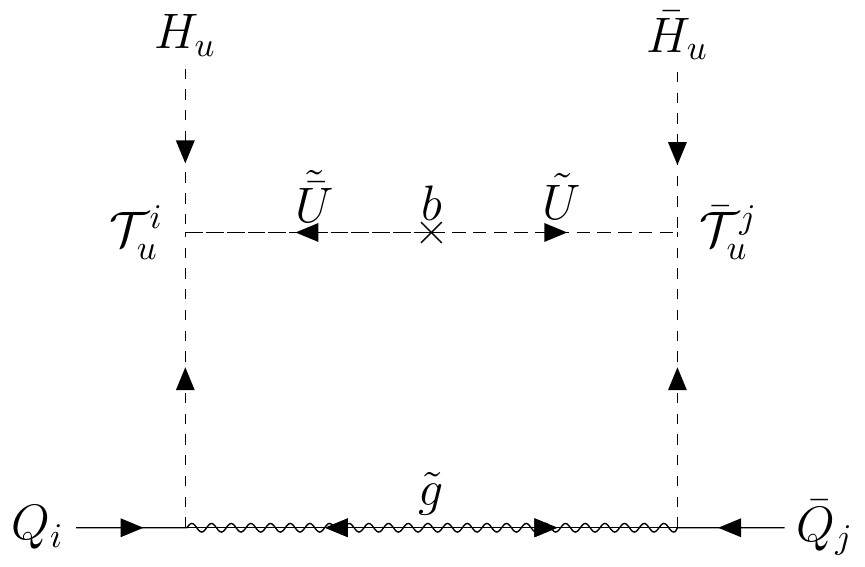}
\end{tabular}	
\end{equation}
and the corresponding contributions, computed for vanishing external momenta,  are
\begin{equation} \label{second}
(\mathcal{M}_u^{\mathcal{T}^2})_{ij}  = \sum_{\kappa=1}^2
\int \frac{d^4 k}{(2\pi)^4}\,\left(2\,C_{F}\,g_3^2\right)\,M_{\tilde{g}}\,P_{\tilde{g}}\,
\mathcal{T}_u^{i}\, P_{\tilde{Q}_{i}} \left(  V^{\dagger}_{1 \kappa}\,P_{\tilde{u}_{\kappa}}\,V_{\kappa 2} \right)
\mathcal{T}_u^{j}\,P_{\tilde{\bar{Q}}_{j}}\, .
\end{equation}

\subsection{Mixed case}

Finally, the simultaneous presence of trilinear terms and mediator mass scale provides a further input to effective Yukawa operator which mixes the previous contributions. As a consequence, the $3\times3$ sub-matrix is given by 
\begin{equation}
\begin{tabular}{m{1cm} m{6cm}}
$(\mathcal{M}_u^{\mu \mathcal{T}})_{ij} \quad =$ &\hspace{1cm} \includegraphics[width=\linewidth]{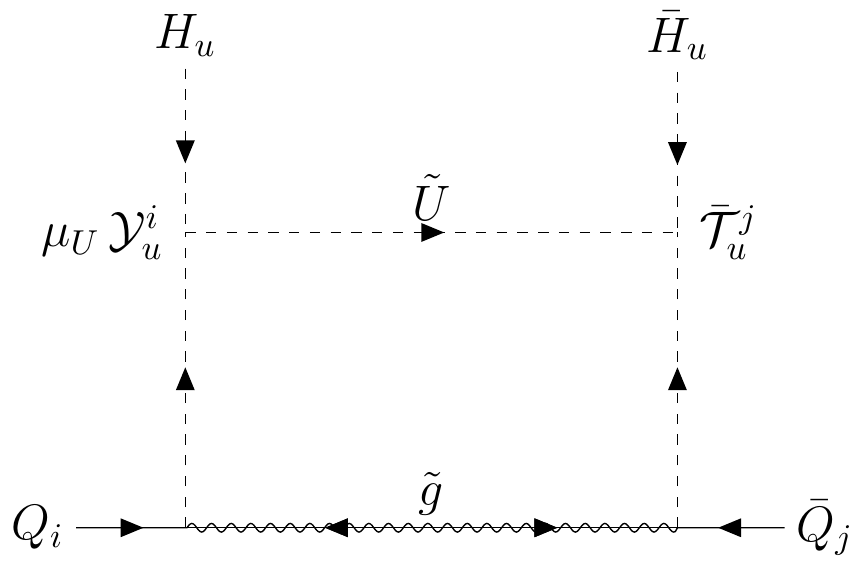}
\end{tabular}	
\end{equation}
plus and additional diagram where the trilinear Yukawa coupling are interchanged. Notice that, differently from the previous cases, the above diagram induces a non-vanishing one-loop contribution to the quark mass matrix even in absence of $b$-terms. Explicitly we have: 
\begin{equation} \label{third}
(\mathcal{M}_u^{\mu \mathcal{T}})_{ij}  = \sum_{\kappa=1}^2
\int \frac{d^4 k}{(2\pi)^4}\,\left(2\,C_{F}\,g_3^2\,\mu_U \right)\,M_{\tilde{g}}\,P_{\tilde{g}}\,
\mathcal{Y}_u^{i}\, P_{\tilde{Q}_{i}} \left(  V^{\dagger}_{2 \kappa}\,P_{\tilde{u}_{\kappa}}\,V_{\kappa 2} \right)
\mathcal{T}_u^{j}\,P_{\tilde{\bar{Q}}_{j}}\, .
\end{equation}

To conclude the section, we remark that the mass matrices of down-type quarks and charged leptons are generated through analogous interactions that involve the corresponding mediators and soft-breaking terms. In particular, replacing all gluino fields in the above diagrams with electroweak gauginos yields additional contributions that are subdominant in the quarks case, but that provide the leading contribution to the radiative masses of charged leptons.

\section{Reproducing the observed mass hierarchy and mixing} 
\label{sec:Or}

\subsection{The quark mass hierarchy}

Challenging eqs.~\eqref{eigM} and ~\eqref{EffYuk}, respectively, with the measured values of the top and charm mass, results in a series of constraints that shape the high energy phenomenology of the \shn{}. 

The top mass sits indeed in the critical range $m_t \sim v_{H}/\sqrt{2}$, forcing the limit $\mathcal Y_u \sim 1$ and $\theta \sim \pi/2$ to recover  $M^2_{u\,3} \simeq \mathcal{K}_{u}^2 \simeq v_{H}^2/2$. Adopting these values in eq~\eqref{eigM} we find that the mediator scale $\mu_U$ is necessarily subdominant with respect to the right-handed breaking scale $v_{\bar H}$. In this case $M^2_{u\,4} \simeq \bar{\mathcal{K}}_{u}^2$ and the model delivers a sharp prediction that relates these scales:
\begin{eqnarray} \label{nobel1}
\frac{M_{u\,4}}{M_{W_R}}\simeq g_2 \sin{\bar{\theta}} \, .
\end{eqnarray} 
Conversely, in the down-quark and lepton sectors, the measured values of the $b$-quark and $\tau$ masses allow for sizable contributions of $\mu_D$ and $\mu_E$ into the corresponding mediator masses.

Assessing the magnitude of these $\mu$ terms at the theory level is crucial for understanding their effect on the loop corrections in eqs.~\eqref{first} and \eqref{third}. Focusing again on the up-type quark case, setting the gluino mass, the squark masses and trilinear terms to a common scale $M_{soft}$ yields the following approximated relations
\begin{eqnarray}
&&\mathcal{M}_u^{\mu^2} \sim \left( C_F \frac{\alpha_S}{\pi} \frac{v_{u} v_{\bar{u}}}{M_{soft}}\right) \frac{\mu_U^2}{M_{soft}^2} \, , \quad 
\mathcal{M}_u^{\mu \mathcal{T}} \sim \left( C_F \frac{\alpha_S}{\pi} \frac{v_{u} v_{\bar{u}}}{M_{soft}}\right) \frac{\mu_U}{M_{soft}}\, , 
\end{eqnarray}
together with the only $\mu$ independent contribution of eq.~\eqref{second} purely determined by the trilinear terms:
\begin{eqnarray}
\mathcal{M}_u^{\mathcal{T}^2} \sim C_F \frac{\alpha_S}{\pi} \frac{v_{u} v_{\bar{u}}}{M_{soft}} \, .
\end{eqnarray}
It is clear that considering $M_{W_R} \gg M_{soft}$ allows to enhance all the considered loop contributions. At the same time, the requirement $\mu_U \ll M_{W_R}$ forced by the experimental value of the top mass and perturbativity implies that the charm mass scale will be inevitably driven by the sole trilinear contributions in $\mathcal{M}_u^{\mathcal{T}^2}$. 

\subsection{The flavor mixing pattern and breaking of flavor symmetry}

On top of generating a suitable quark mass spectrum, to explain the observed flavor structure we must address the challenging problem posed by the CKM mixing matrix. This is characterized by an almost symmetric structure, with sizable mixing only in the Cabibbo sub-matrix acting on the lightest quark generations. 

To investigate this mixing pattern within the \shn{}, we simplify the matter by considering up and down Yukawa vectors oriented in flavor space along the direction corresponding to the third generation quarks: $\mathcal{Y}_i = (0,0,\mathcal{Y})$. Then, from the discussion in Sec.~\ref{sec:Ofcb}, it is clear that any component of the trilinear terms parallel to the Yukawa direction would only result in radiative corrections to the tree-level top and bottom masses. To avoid this unnecessary complication, we consider the ansatz 
\begin{eqnarray}
\mathcal{T}^i_u = \mathcal{T}_u (\cos \alpha_u, \sin \alpha_u, 0)\,,\,\,\, \,\,\mathcal{T}^i_d = \mathcal{T}_d (\cos \alpha_d, \sin \alpha_d, 0)\,, 
\end{eqnarray}  
which, supplemented with non-degenerate diagonal soft squark mass terms, breaks the residual $U(2)$ flavor symmetry present at the tree-level in the model.  

Notice that the conditions at the basis of the large mixing of the light quarks in the up-type sector are a direct consequence of the measured value of the top mass, which bounds the magnitude of $\mu_U$. Indeed, considering our ansatz and the properties of the loop functions in Sec.~\ref{sec:Ofcb}, a limited $\mu_U$ ensures a natural separation between the third generation quarks and the light states in the $U(2)$ flavor subspace, that obtain their masses through the contributions of trilinear terms.     

Motivated by this observation, in the present work we shape the mixing of the down sector in a similar fashion by imposing an analogous ansatz on the $\mu_D$ term: 
\begin{eqnarray} \label{nobel2}
  \frac{M_{d\,4}}{M_{W_R}}\simeq g_2 \mathcal{Y}_d \cos{\bar{\theta}} \,.
  \end{eqnarray}  
With this choice we effectively ascribe the observed CKM mixing to a mild mismatch between the matrices that perform the rotation between mass and flavor eigenstates in the up and a down-type sectors. We find this approach promising because both the matrices are already characterized by a large mixing amongst the first and second generation, although we do not exclude that solutions of the model which match the measured CKM entries for more general down-type mixing matrix could exist.

\section{Numerical analysis} 
\label{sec:FH}

Explaining the observed flavor structure in terms of dynamics within a natural theory is a highly non-trivial challenge. The apparent lack of a principle behind the measured span of masses and mixing angles provides no guidance for an understanding of the underlying structure. On top of that, collider experiments have now pushed the most popular models of new physics well above the TeV scale, limiting their possible contributions to the flavor puzzle. 

A first concrete obstacle on our path is the difference in the mass hierarchies measured in the up- and down-type sector where, respectively, $m_{u}/m_{t}\sim10^{-5}$ and $m_{d}/m_{b}\sim10^{-3}$, and the $d$-quark is heavier than the $u$-quark. 

A second problem is posed by the measured CKM elements
\begin{equation}
|V_{CKM}| = \left[
\begin{array}{ccc}
0.97485 \pm 0.00075 & 0.2225 \pm 0.0035 & 0.00365 \pm 0.00115 \\ 
0.2225 \pm 0.0035  & 0.9740 \pm 0.0008 & 0.041 \pm 0.003 \\
0.0009 \pm 0.005 & 0.0405 \pm 0.0035 & 0.99915 \pm 0.00015
\end{array}
\right]\, ,
\end{equation}
which result in an almost symmetric structure spoiled only by the entries corresponding to top-down and up-bottom transitions. Furthermore,  mixing is sizable only in correspondence of the Cabibbo sub-matrix\footnote{For the purposes of this analysis we neglect the measured $CP$-violating phase.}.

Regardless of all these difficulties, we will now show that the \shn{} is able to faithfully reproduce the flavor texture of the SM, matching both the hierarchies of the quark sectors as well as their mixing pattern. The new particles that the model predicts are all well above the TeV scale, even when perturbative values of the couplings are imposed. 

\subsection{Methodology of the scan}

Our investigation comprises two different parts: the first concerns a scan on  supersymmetry preserving parameters, whereas the second assesses the effect of varying soft terms. As a first step,  we use in eq.~\eqref{eigMEx} values of top and bottom quark masses taken from the corresponding 68\% confidence intervals, implicitly neglecting possible radiative corrections to these quantities. Although we adopt such a simplification for this first numerical check of the model, these contributions can be straightforwardly included in future analyses. 

As a second step, we scan on $M_{u\,4}$, $M_{d\,4}$ and $M_{W_R}$ in the range shown in eq.~\eqref{tab:SS}, motivated by our previous considerations on the relevance of the ratio $v_{\bar H}/M_{soft}$. In order to avoid large radiative corrections to the third generation masses, as well to accommodate a sizable mixing between the light generations, we then retain only points with $\mu_D \leq 5$ TeV, with the exception of the analysis presented in the panels of Fig.~\ref{fig:11} where the extreme cases of $\mu_D \leq 1 $ TeV and large $\mu_D$ are investigated. 
The condition $\mu_D \leq 5$ TeV then implements the limit in eq.~\eqref{nobel2} on top of the one in eq.~\eqref{nobel1} due to the measured value of the top mass,
\begin{equation} \label{tab:SS}
\begin{array}{ccc}
M_{u\,4} \in [5,10]\text{ TeV}\,,  & 
M_{d\,4} \in [5,20]\text{ TeV}\,,& 
M_{W_R} \in [5,10]\text{ TeV}\,.
\end{array}
\end{equation}

At this stage we filter the collected points requiring perturbative values of the Yukawa couplings and real $\mu_{U/D}$ terms. The selected configurations are then used to initiate the second scan that targets the soft breaking parameters, which are varied according to the ranges in eq.~\eqref{tab:SB} 
\begin{eqnarray} \label{tab:SB}
&&M_{soft} \in [5,10]\text{ TeV}\,\,,\,\,  
\Delta M_{soft} = M_{soft}/4\text{ TeV} 	 \,,  \nn\\ 
&&m_{Q}^{1,1} = m_{\bar{Q}}^{1,1} = 1\text{ TeV} \,,\nn\\ 
&&m_{Q}^{2,2} = m_{\bar{Q}}^{2,2} = m_{Q}^{3,3} = m_{\bar{Q}}^{3,3}\in [M_{soft},M_{soft}+ \Delta M_{soft}]  \,, \nn\\ 
&& |\mathcal{T}_u| \in [M_{soft},M_{soft}+ \Delta M_{soft}]\, , \,\,|\mathcal{T}_d| \in [M_{soft},M_{soft}+ \Delta M_{soft}] \,,\nn\\ 
&& m_{\tilde{u}_1} \in [M_{soft},M_{soft}+ \Delta M_{soft}]\, , \,\, m_{\tilde{u}_2} - m_{\tilde{u}_1} \in [-\Delta M_{soft},\Delta M_{soft}] \,,\nn \\  
&& \alpha_u \in [0,\pi/8]\, , \,\, \alpha_d \in [\pi/4,\pi/2] \,,\nn \\
&& M_{\tilde g}  \in [M_{soft},M_{soft}+ \Delta M_{soft}]\,.
\end{eqnarray}
Notice that we indicated here with $m^2_{\tilde{u}_i}$ the (s)messengers masses above the left-right symmetry breaking scales, cf. appendix \ref{bterms}. To conclude, we remark that the ansatz adopted allows to recognize the parameters responsible for the generation of the masses of light quarks. These are the component ($\alpha_u$ and $\alpha_d$) of the trilinear terms orthogonal in flavour space to $\mathcal Y_{u/d}$ and the non degeneracy between the first-generation and the remaining two squark soft masses.

\subsection{Avoiding instabilities and colour breaking}
Before discussing the results obtained with the above methodology, we need to face a further potential problem rooted in the necessary use of trilinear terms for reproducing the up-type quark mass spectrum.

The presence of sizable trilinear terms, in fact, could source instabilities in the scalar potential and even trigger the appearance of colour-breaking vacuum states. We therefore need to check that at the minimum of the potential, where spontaneous symmetry breaking of the left-right symmetry is achieved, the soft squark mass terms and the trilinear used in the analysis do not lead to tachyonic instabilities. The problem then reduces to the computation of the (s)messanger and squarks full mass matrix eigenvalues on the left-right breaking vacuum.  

A further source of instabilities is due to possible high energy configurations of the fields, which may induce electric or colour charge-breaking minima. We investigate this possibility by considering a direction where all the scalar components of the fields participating in the Yukawa interactions develop the same vacuum expectation value. For instance, leaving gauge and generation indices understood, for the case of up-type quarks we have
\begin{eqnarray}
|\tilde{\bar{U}}|=|\tilde{U}|=|\tilde{\bar{Q}}|=|\tilde{Q}|=|\tilde{\bar{H}}_u|=|\tilde{H}_u| = \frac{\phi}{\sqrt 6} \,,
\end{eqnarray}  
subject to the scalar potential
\begin{eqnarray} \label{potBorzu}
V(\phi) = \frac{S^2}{2} \phi^2 + \frac{T}{3\sqrt{6}} \phi^3 + \frac{L^2}{36}\phi^4 \, . 
\end{eqnarray} 
In a supersymmetric model, the $d$-terms and quartic Yukawa interactions enter the definition of the coefficient $L$, while soft trilinear terms and superpotential induced $f$-terms shape the T coefficient. Quadratic soft terms and Dirac masses define, finally, the coefficient S. In order to avoid the breaking of the color and charge symmetries, the trilinear terms must satisfy the condition
\begin{eqnarray} \label{borzu}
\frac{T}{S} > - \sqrt{3} L \,,
\end{eqnarray} 
derived from eq.~\eqref{potBorzu} under the requirement that non trivial minima  have energies larger than the trivial one corresponding to $\phi=0$.

Unfortunately, the implementation of the above condition in our scan would require a detailed study of the Higgs sector of the model and of the high energy evolution of quantum corrections to eq.~\eqref{potBorzu} at the scale $\phi$. For the purpose of the present analysis, however, it is sufficient to notice that our fit of the lightest quark and lepton masses will hardly worsened the condition in eq.~\eqref{borzu} as it generally involves trilinear terms and soft masses of the same order. Furthermore, arbitrary large values of $m_Q^{3,3}$ and $m_{\bar Q}^{3,3}$ can be used here to stabilize the potential without consequences for the radiative masses generated.

\subsection{Results}
 
In this section we finally gather the explicit results obtained by the \shn{} for the mass hierarchy and mixing of quarks. 

\begin{figure}[ht!]
	\centering
	\includegraphics[width=0.45\linewidth]{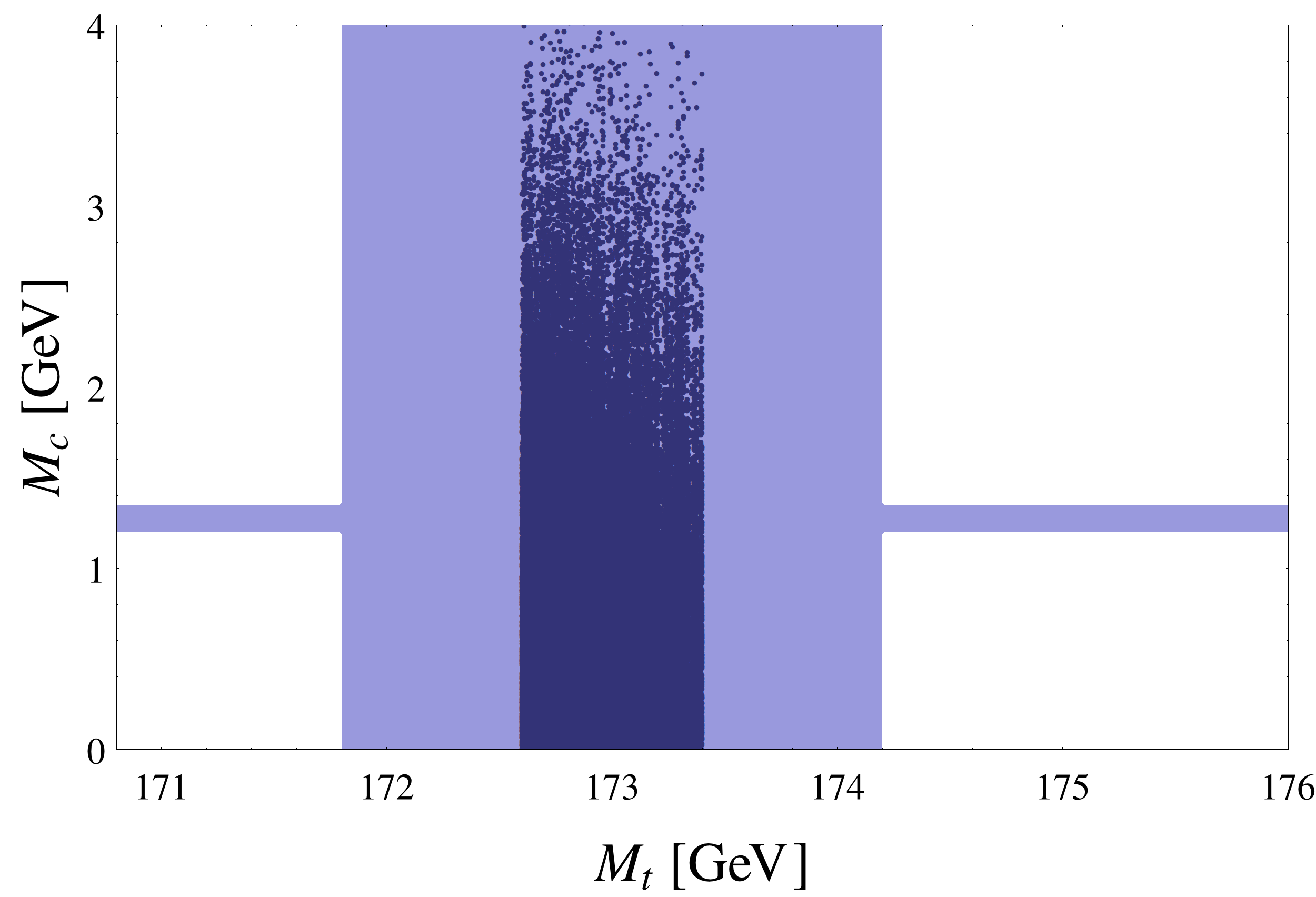}
	\includegraphics[width=0.45\linewidth]{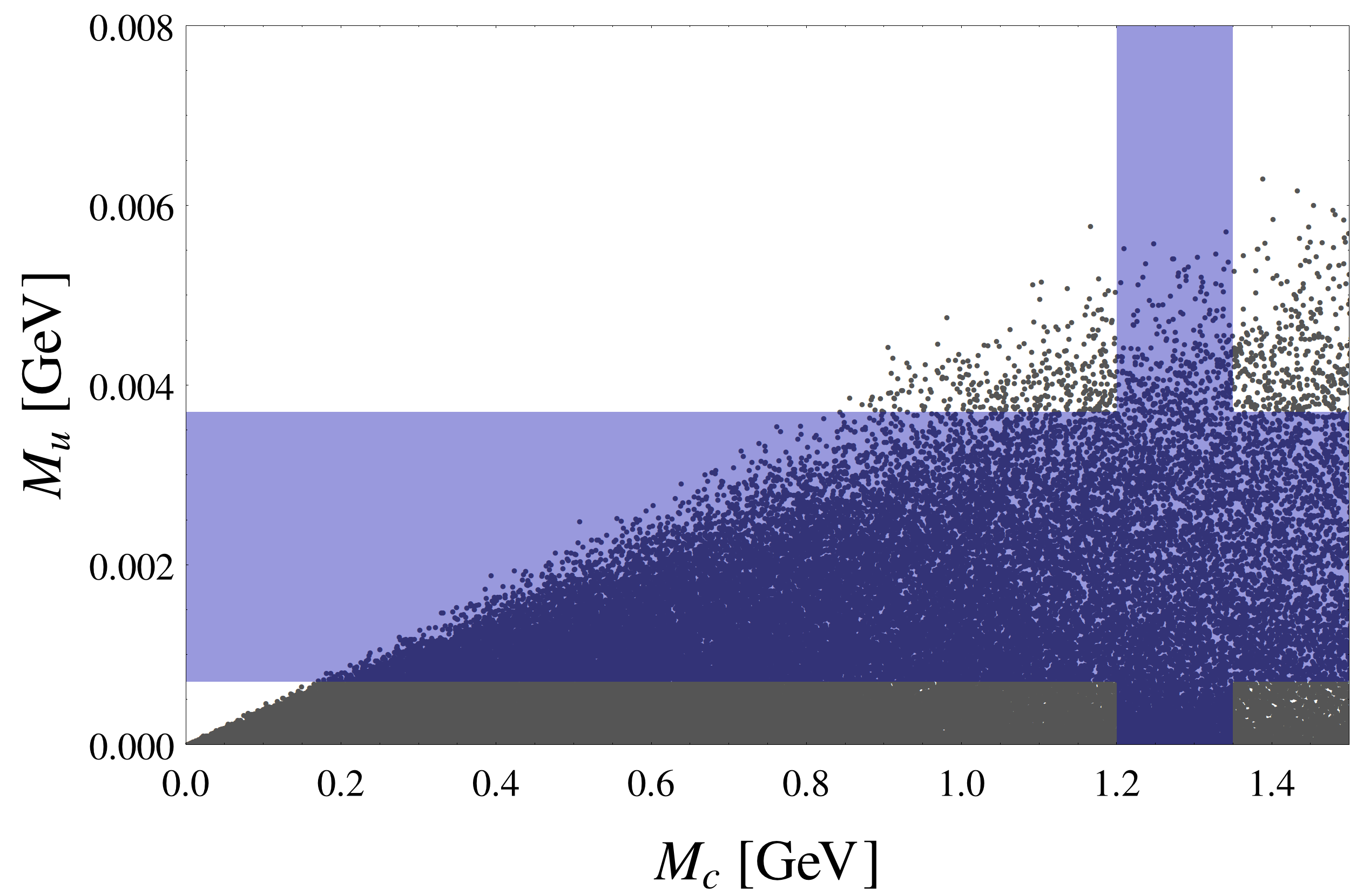} 
	\caption{Predictions of the model for the up-type quark masses. Left panel: $M_t$ and $M_c$. Right panel: $M_c$ and $M_u$. The blue bands indicate the  3$\sigma$ confidence intervals for the measured values of the masses.}
	\label{fig:10}
\end{figure}

Starting with the up-type case presented in Fig.~\ref{fig:10}, the left panel shows that the model can easily reproduce the hierarchy in mass between top and charm quarks. The tree-level and one-loop generation mechanisms of the effective Yukawa operators are essentially uncorrelated and, as a consequence, it is possible to simultaneously reach well within the experimental $3\sigma$ confidence interval of both quarks indicated by the light blue bands. This behaviour is due to our ansatz for the trilinear terms, which effectively decouples the tree-level dynamics encoded in the messenger Yukawa from the radiative generation of charm and up quarks. We therefore expect that new correlation between $M_t$ and $M_c$ appear in more general analyses where the trilinear term posses a component parallel, in flavour space, to the Yukawa vector that identifies the third SM generation. In spite of that, the present result are indeed sufficient to demonstrate that the \shn{} is able to achieve the required hierarchy.

As for the masses of the first two generation quarks, in the right panel we see the presence of a correlation that favours lower values of $M_u$ once a central value for $M_c$ is considered. Such a behaviour is induced by the trilinear terms, which control here the radiative generation of both the masses, and by the contribution of non-degenerate squark masses that are essential for obtaining a non vanishing value of the up quark mass. We stress that in this first investigation of the model we have neglected effects due to the renormalization group running of the relevant effective operators between the matching and the considered mass scale, which could have a sizable impact on the predictions for the first generation quark masses.

\begin{figure}[h]
	\centering
	\includegraphics[width=0.45\linewidth]{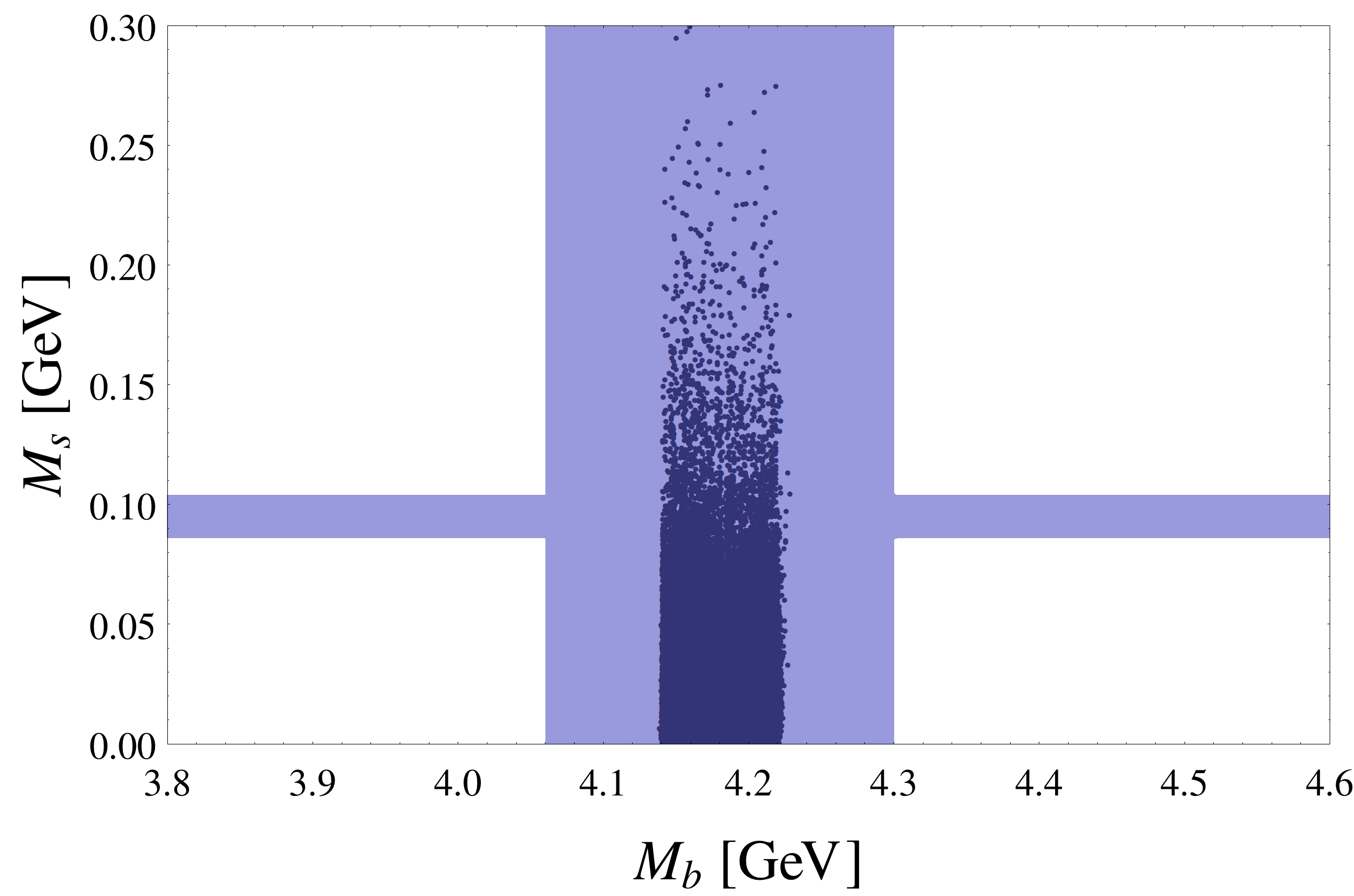} 
	\includegraphics[width=0.45\linewidth]{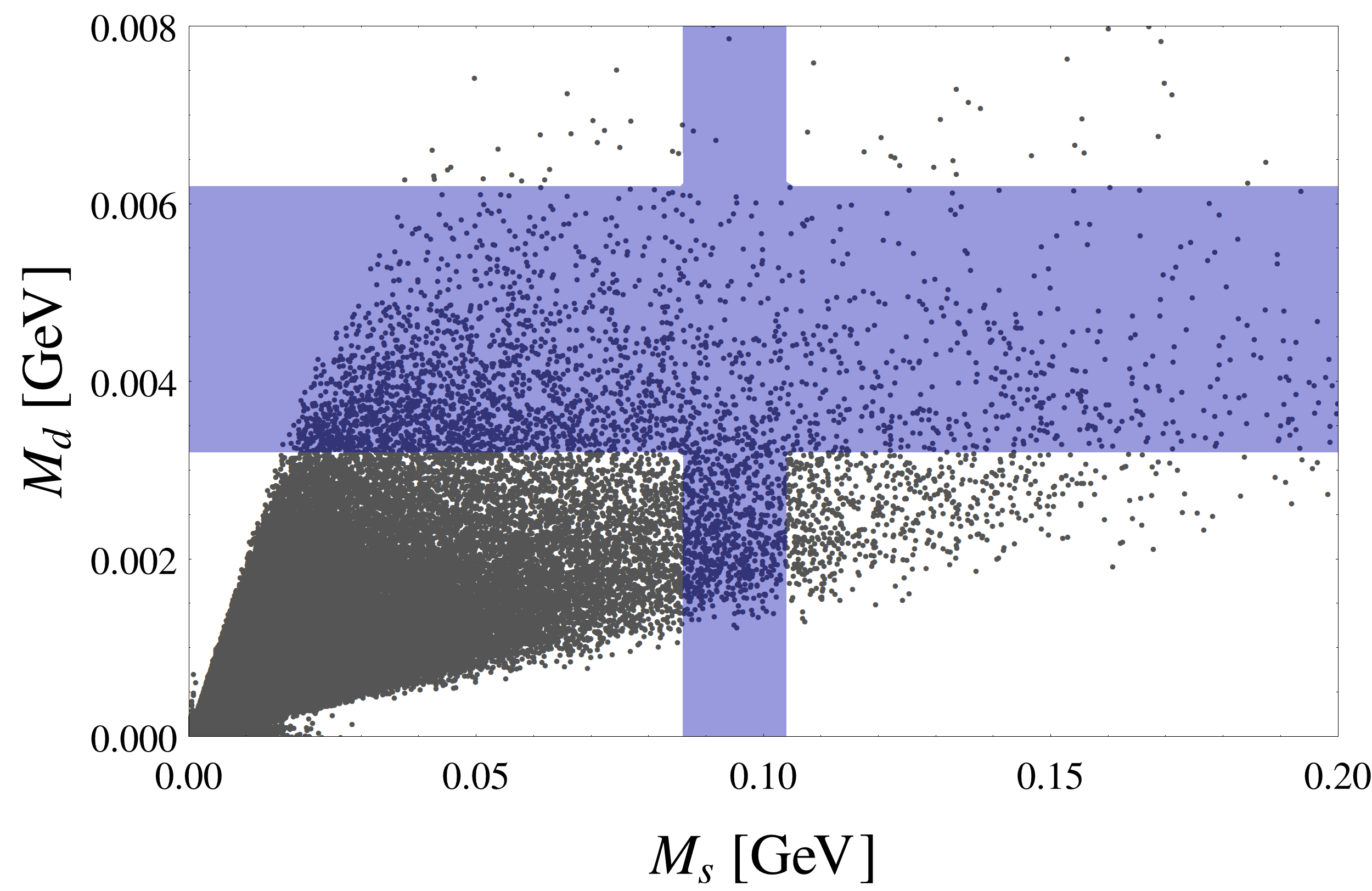} 
	\caption{Predictions of the model for the down-type quark masses. Left panel: $M_b$ and $M_s$. Right panel: $M_s$ and $M_d$. The blue bands indicate the  3$\sigma$ confidence intervals for the measured values of the masses.}
	\label{fig:11}
\end{figure}

Our results for the down sector, shown in Fig.~\ref{fig:11}, are qualitatively the same. For our choice of trilinear terms, the solutions of the model presented in the left panel do not correlate the bottom quark mass with that of the strange quark, allowing to reach an excellent agreement with the current measurements. Differently, barring possible renormalization group evolution contributions, we see that the model tends to prefer smaller values of both $M_d$ and $M_s$, although good agreement with the data is still achievable.   

\begin{figure}[h]
	\centering
	\includegraphics[width=0.42\linewidth]{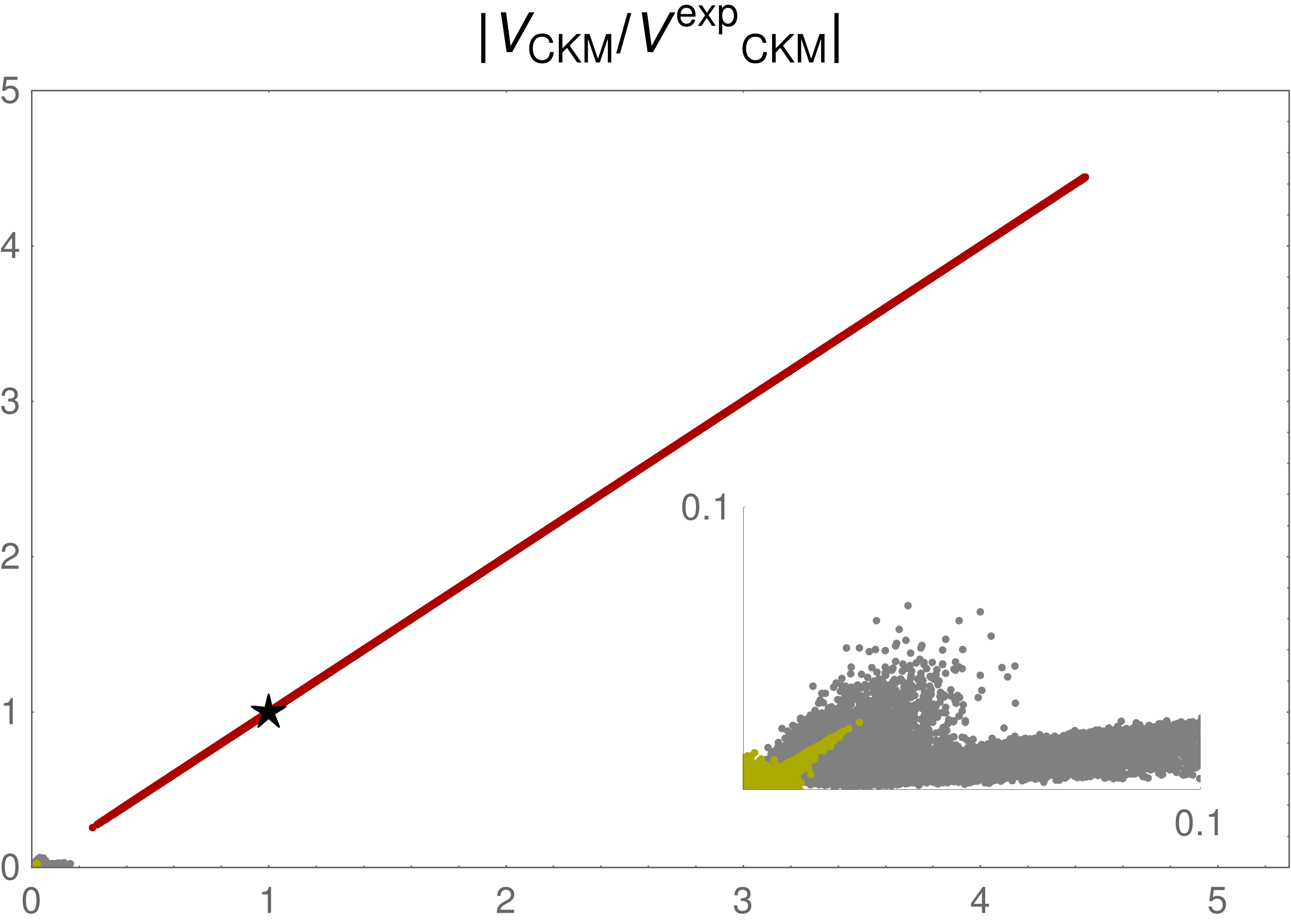}
	\includegraphics[width=0.42\linewidth]{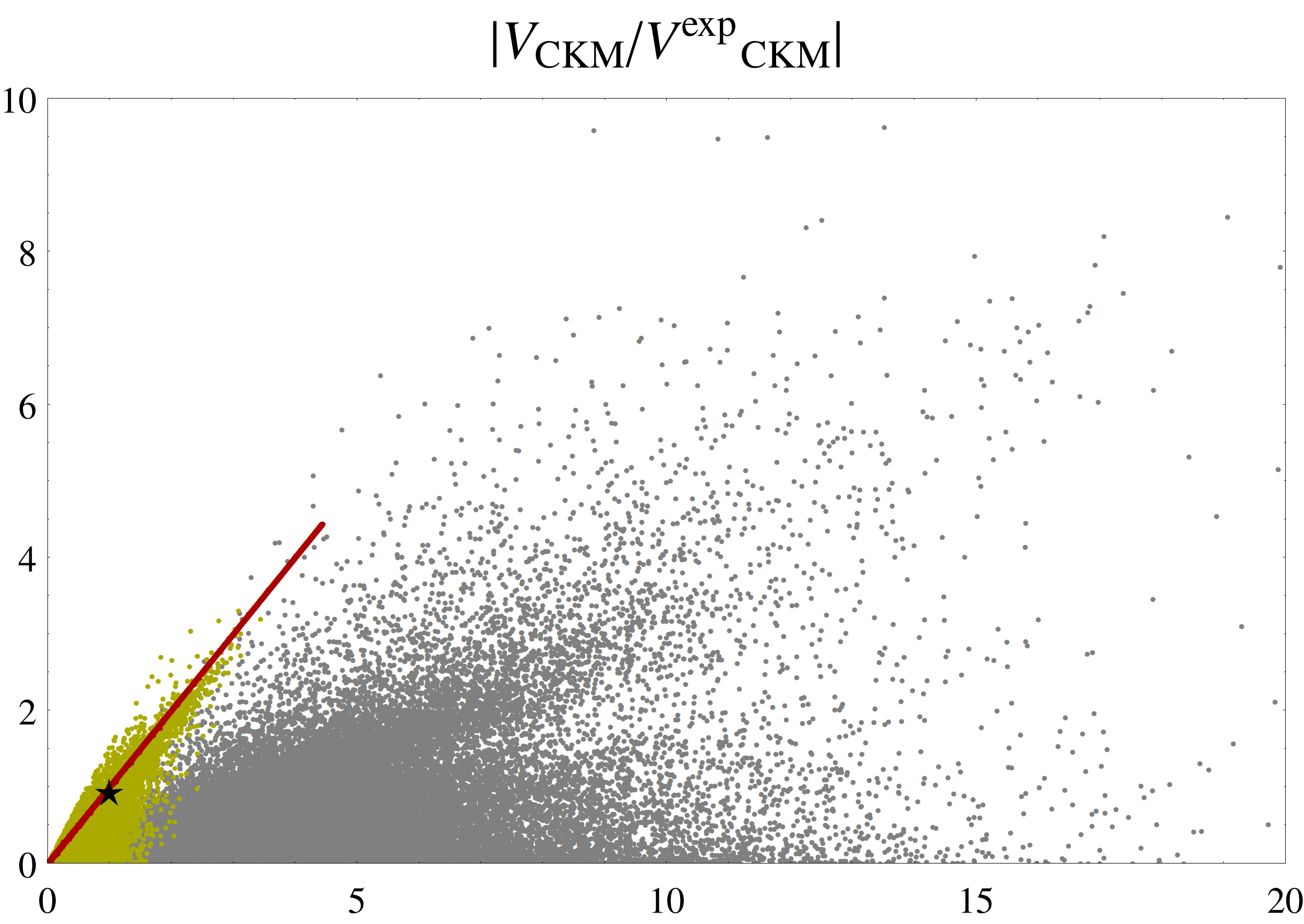}
	\caption{Predictions of the model for the elements of the CKM mixing matrix normalized with respect to the corresponding experimental best-fit values. The red points correspond to the $\{V_{1\,2},V_{2\,1}\}$ elements, which regulate the Cabibbo sector, whereas the gray and yellow point indicate $\{V_{1\,3},V_{3\,1}\}$  and $\{V_{3\,2},V_{2\,3}\}$, respectively. In the left panel, $\mu_D$ is set in the same range as $\mu_U$ ($\mu_D \sim \mu_U \leq 1$). In the right panel, instead, $\mu_D$ ranges on all the values allowed by eq.~\eqref{rel2}, explicitly: $ \mu_D \leq M_{u\,4}$. }
	\label{fig:12}
\end{figure}

Finally, regarding the quark mixing, we plot in Fig.~\ref{fig:12} the predictions of the model for the CKM matrix elements relative to the corresponding experimental best fit values. In the left panel we limit the $\mu_D$ term to below a few TeV, whereas in the right panel we let it vary. In both plots we indicate with red points the solutions obtained in the \shn{} for the Cabibbo mixing, regulated by the $\{V_{12}, v_{21}\}$ elements of the CKM matrix. Analogously, the yellow and grey points represent the value obtained for the $\{V_{32}, v_{23}\}$ and $\{V_{13}, v_{31}\}$ elements, respectively. All quantities have been rescaled by the corresponding best fit measurements, matched in correspondence of the (1,1) point indicated by a star. As we can see, the model straightforwardly reproduces the observed Cabibbo mixing regardless of the maximal amplitude of $\mu_D$. Interestingly, we find instead that matching the magnitude of the remaining mixing angles requires larger values of this parameter, pointing to the presence of a mild hierarchy in the mediator sector. This is not a concern for the \shn{}, as a precise determination of the CKM elements barring the Cabibbo sector would require a careful assessment of the renormalization group evolution contribution that we neglect in this first analysis. 

\FloatBarrier

\section{Conclusions and outlook} 
\label{sec:Conclusions}
In this paper we introduced the \shn{}, a new framework based on the interplay between two well-know symmetries: the supersymmetry and the left-right gauge symmetry. 

By employing only doublet representations for the scalar sector, the left-right symmetry forbids the usual four-dimensional Yukawa operator and consequently lays the foundation for their effective generation. In the proposed model we considered a minimal set of three new mediator (super)fields, which possess color and $B-L$ charge and that connect the two chiral sectors of the theory. As a consequence, below the left-right breaking scale, the charged fermions of the third SM generation acquire tree-level effective Yukawa interactions that reproduce the SM couplings. 

Supersymmetry, instead, provides the missing ingredients for the radiative generation of the remaining Yukawa couplings. This can be achieved only at the loop-level, thereby explaining the lightness of the first two SM fermion generations in a natural way. The Majorana gaugino masses, soft trilinear terms and squark masses, respectively, operate the breaking of any remnant chiral and flavour symmetry, ensuring the emergence of mass terms for these particles. On general grounds, from the perspective of low-energy SUSY, the appearance of hierarchical fermion masses and mixing is rather unexpected and, according to conventional wisdom, SUSY should not shed any light on the problem. Instead, we demonstrated here that the well-known minimal flavour violation~\cite{DAmbrosio:2002vsn} ansatz, usually imposed on SUSY flavour models by hand, may actually be a natural consequence of the framework due to the {\it effective} nature of the Yukawa couplings.

As a first test for the \shn{}, in this paper we faced the problem of generating the SM flavor structures. Our findings show that the model straightforwardly matches the SM quark spectrum considering a new sector with mediators and right-handed Higgs boson at a few TeV scale, and internal hierarchies not larger than one order of magnitude. Remarkably, we also find that the model easily recovers the observed Cabibbo mixing in the same parameter space. Matching the remaining CKM elements, instead, seemingly requires a mild hierarchy in the mediator sector. We however remark that a careful investigation of these quantities, as well as of the lightest quark masses, should properly account for renormalization group evolution effects which we have neglected in this first analysis.    

Beside proposing a new solution to the flavour puzzle that falls necessarily within the reach of next-generation collider experiments, the \shn{} phenomenology remains to-date largely unexplored.  The presence of four Higgs boson doublets, for instance, could have implications on the present B-mesons flavour anomalies~\cite{Marzo:2019ldg} and allow for novel collider signatures of the framework. Likewise, the peculiar neutrino mass generation mechanism, which takes place at the two-loop level, predicts Dirac neutrinos necessarily lighter than the remaining fermions and could be at the basis of the different mixing observed in the quark and lepton sector. Lastly, because within the \shn{} all fermion Yukawa couplings cease to be local operators above the $SU(2)_R$ breaking scale, the framework predicts important modifications to the standard renormalization group evolution of the remaining quantities which may substantially impact the dynamics of grand-unified scenarios.

\section*{Acknowledgments}
This work was supported by the Estonian Research Council grants IUT23-6, PRG356 and by the EU through the ERDF CoE program project TK133. EG would like to thank the CERN Theory Department for its kind hospitality during the preparation of this work.

\section*{Appendices:} 

\begin{appendices}

We report below the explicit expressions, vertices and Feynman rules that enter the computation of the quark mass matrix in the present framework.   

\section{Feynman rules for up sector} \label{FRUp}

The terms relevant to the performed diagrammatic computations can be isolated from eq.~\eqref{superpot} and \eqref{supersoft} after solving for the auxiliary fields
\begin{eqnarray} \label{RelevantLag}
\mathcal{L}' &=&  - \left(\sqrt{2}\, g_3\right)\,\left(\tilde{Q}^{\dagger}_{i} T^a Q_{i}\, \tilde{g} 
+ \,\tilde{\bar{Q}}^{\dagger}_{i} \bar{T}^a \bar{Q}_{i}\, \tilde{g}  \right)  - \mathcal{Y}^{i} Q_i\,H_u \,\bar{U} + \bar{\mathcal{Y}}^{i} \,\bar{Q}_{i} \bar{H}_u \,U - \mu_U U \bar U  + \nn \\
&&  - |\mu_U|^2 \tilde{U}^* \tilde{U} - |\mu_U|^2 \tilde{\bar{U}}^* \tilde{\bar{U}} - \mu_U^* \mathcal{Y}^{i} \tilde{Q}_i\,H_u \,\tilde{U}^* + \mu_U^* \bar{\mathcal{Y}}^{i} \tilde{\bar{Q}}_i\,\bar{H}_u \,\tilde{\bar{U}}^*
+ b_{\mu_U} \tilde{U} \tilde{\bar U} + \text{c.c.}  + \mathcal{L}_{Soft}\,, \nn \\ 
\end{eqnarray}
where $\bar{T}^a$ are the generators of $SU(3)$ in the anti-fundamental representation. The remaining neutral gaugino interactions involving coloured states have been neglected because subdominant with respect to the gluino exchange. 

\subsection{Mixed and diagonal propagation for scalars}
\begin{figure}[ht!]
	\centering
  \includegraphics[scale=0.6]{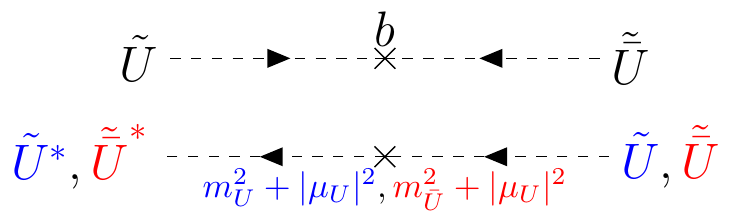}
	\includegraphics[scale=0.6]{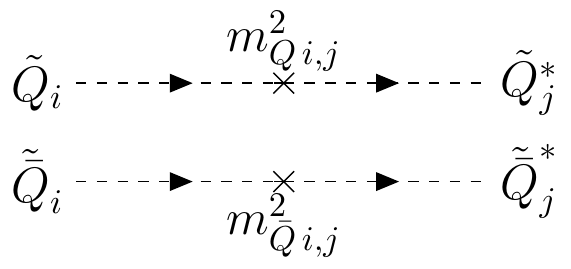}
	\caption{Clockwise starting from bottom left: Feynman rules for diagonal smessengers propagation, smessengers mixing, left and right-handed squarks mixing.}
	\label{fig:0}
\end{figure}

\subsection{Vertices}

\begin{figure}[ht!]
\centering
\includegraphics[scale=0.7]{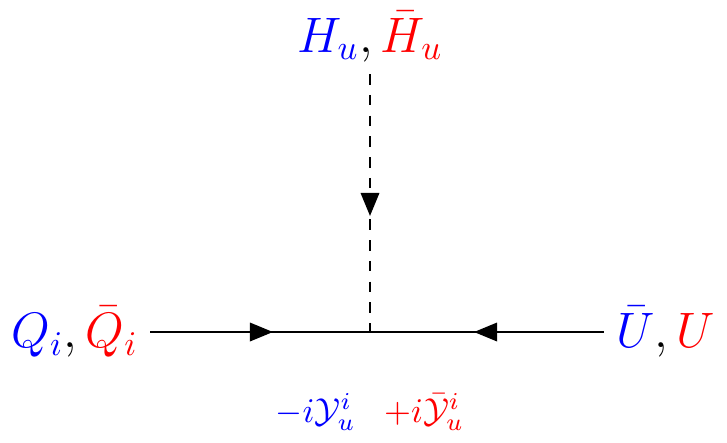}
\caption{Higgs-Quark-Messenger interaction, left-handed and right-handed cases.}
\label{fig:1}
\end{figure}
\vspace{0.5 cm}
\begin{figure}[ht!]
\centering
\includegraphics[scale=0.6]{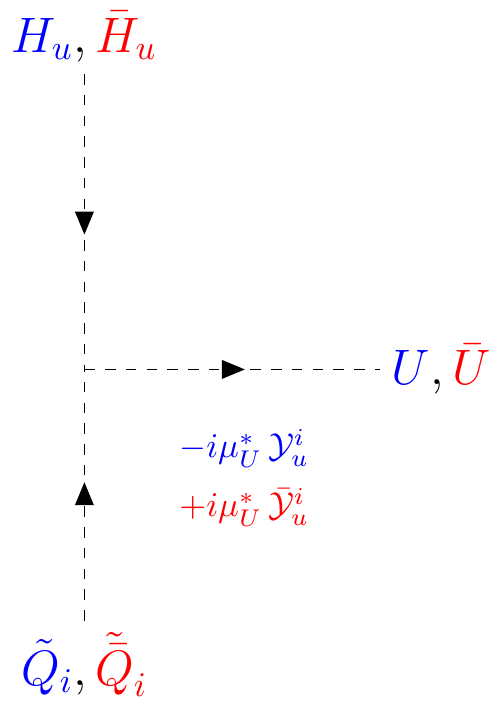}
\includegraphics[scale=0.6]{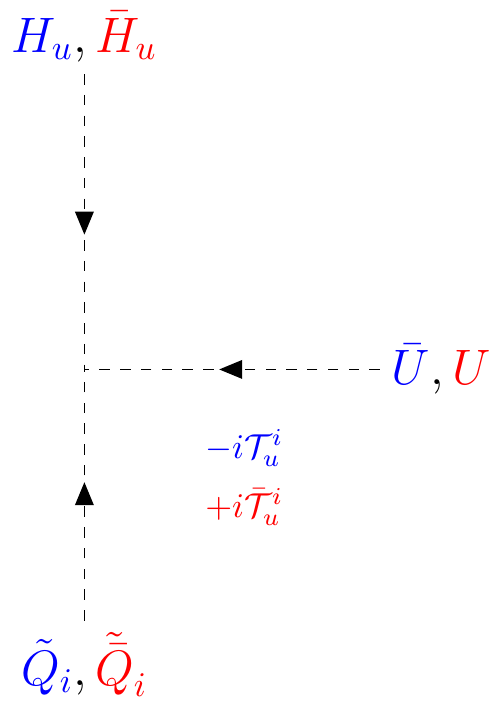}
\includegraphics[scale=0.6]{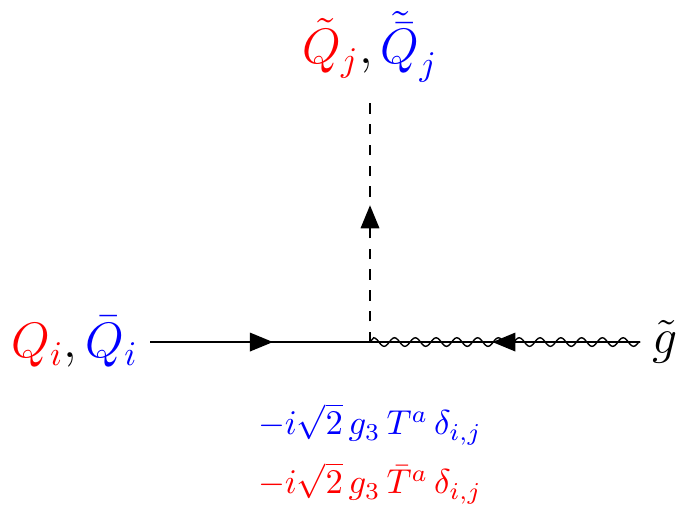}
\caption{Squark-Higgs-Scalar messenger vertices and gluino vertex.}
\label{fig:2}
\end{figure}

\FloatBarrier

\subsection{Full one-loop gluino contribution} \label{fullLoop1}

\begin{equation} \label{fullLoop2}
(\mathcal{M}_u^{\mu \mathcal{T}})_{ij}  = \sum_{k=1}^2 \frac{2 \alpha_S}{3 \pi} M_{\tilde{g}}  \left(\mu_U \mathcal{Y}_u^i V^{\dagger}_{2,k} +\mathcal{T}_u^i V^{\dagger}_{1,k}  \right)D_0 [M_{\tilde g}, m_{Q_i}, m_{\bar{Q}_j}, m_{u_k}] \left(\mu_U^* \mathcal{Y}_u^j V_{k,1} +\mathcal{T}_u^j V_{k,2}  \right)\,.
\end{equation}

\subsection{Zero momentum 4-point loop function}
The zero-momentum $D_0$ used in eq.~\eqref{fullLoop1} is given by 
\begin{eqnarray}
&&D_0\left[m_1,m_2,m_3,m_4\right] = \int \frac{d^4 k}{i\,\pi^2}\,\frac{1}{k^2 - m_1^2}\frac{1}{k^2 - m_2^2}\frac{1}{k^2 - m_3^2}\frac{1}{k^2 - m_4^2} = \nn \\
&&\left(-\frac{m_1^2 \log \left(\frac{m_1^2}{m_3^2}\right)}{\left(m_1^2-m_2^2\right) \left(m_1^2-m_3^2\right)
	\left(m_1^2-m_4^2\right)}+\frac{m_2^2 \log \left(\frac{m_2^2}{m_3^2}\right)}{\left(m_1^2-m_2^2\right) \left(m_2^2-m_3^2\right)
	\left(m_2^2-m_4^2\right)} \right. \nn \\
&&\left. +\frac{m_4^2 \log \left(\frac{m_3^2}{m_4^2}\right)}{\left(m_4^2-m_1^2\right) \left(m_4^2-m_2^2\right)
	\left(m_4^2-m_3^2\right)}\right)\,,
\end{eqnarray}
which is further reducible if at least two of the masses in the loop are equal.

\subsection{B-term and (s)messenger mixing} \label{bterms}
The complex scalar fields $\tilde U$ and $\tilde{\bar U}$ mix according to
\begin{eqnarray}
\left(\begin{array}{cc}
	\tilde{U}^* & \tilde{\bar{U}}
\end{array}
\right) \left(
\begin{array}{cc}
	m^2_U + \mu_U^2 & b \\
	b & m^2_{\bar{U}} + \mu_U^2
\end{array}
\right) \left(\begin{array}{c}
	\tilde{U} \\ \tilde{\bar{U}}^*
\end{array}
\right) = 
\left(\begin{array}{cc}
	\tilde{u}_{1}^* & \tilde{u}^*_2
\end{array}
\right) \left(
\begin{array}{cc}
	m^2_{\tilde{u}_1} & 0 \\
	0 & m^2_{\tilde{u}_2}
\end{array}
\right) \left(\begin{array}{c}
	\tilde{u}_{1} \\ \tilde{u}_2
\end{array}
\right)\,.
\end{eqnarray}
It is clear from Fig.(\ref{diags}-\ref{diags2}) that a non-vanishing $b$-term is essential for the diagram to have a non-zero contribution. If we consider the case $m^2_{U} = m^2_{\bar{U}}$, before the left-right symmetry breaking, the two massive combinations
\begin{eqnarray}
\tilde{u}_1 &=& V_{1,1}\,\tilde{U} + V_{1,2}\,\tilde{\bar{U}}^*  =\frac{1}{\sqrt{2}} \left(-\tilde{U} + \tilde{\bar{U}}^* \right)\,,\,\, m^2_{\tilde{u}_1} = (m^2_U + \mu_U^2 - b)\,,\nn \\
\tilde{u}_2 &=& V_{2,1}\,\tilde{U} + V_{2,2}\,\tilde{\bar{U}}^* =\frac{1}{\sqrt{2}} \left(\tilde{U} + \tilde{\bar{U}}^* \right)\,,\,\, m^2_{\tilde{u}_2} = (m^2_U + \mu_U^2 + b)\, ,
\label{smesMas}
\end{eqnarray}
propagate in the loops. 
In turn, eq.~\eqref{smesMas} then sets the magnitude of $m^2_U$ and $b$ as
\begin{eqnarray}
b = \frac{ m^2_{\tilde{u}_2}- m^2_{\tilde{u}_1}}{2} \,, m^2_U = \frac{ m^2_{\tilde{u}_2}+ m^2_{\tilde{u}_1}}{2} - \mu_U^2\,.
\end{eqnarray}

\end{appendices}



\end{document}